\documentclass[twoside]{IEEEtran}

\usepackage{amsfonts}
\usepackage[dvips]{graphicx}
\usepackage{times}
\usepackage{cite}
\usepackage{amsmath}
\usepackage{color}
\usepackage{cases}
\usepackage{array}
\usepackage{dsfont}
\usepackage{amssymb}

\usepackage{stfloats}
\usepackage{slashbox}
\usepackage{graphicx}
\usepackage{footnote}
\usepackage{booktabs}
\usepackage{array}
\usepackage{bm}
\usepackage{multicol}
\usepackage{balance}

\begin{document}

\title{Generalized Signal Alignment: On the Achievable DoF for Multi-User MIMO Two-Way Relay Channels}
\author{\IEEEauthorblockN{Kangqi Liu, \IEEEmembership{Student Member,~IEEE} and Meixia Tao, \IEEEmembership{Senior Member,~IEEE}}\\
\thanks{Manuscript received May 4, 2014; revised December 5, 2014 and March 30, 2015; accepted April 1, 2015. This work was presented in part at the 2014 IEEE International Conference on Communications \cite{Liu3}, the 2014 IEEE Global Telecommunications Conference \cite{Liu5} and the 2015 IEEE International Conference on Communications \cite{Liu6}. The authors are with the Department of Electronic Engineering, Shanghai Jiao Tong University, Shanghai, China (Emails: k.liu.cn@ieee.org, mxtao@sjtu.edu.cn). This work is supported by the National 973 project of China under grant 2012CB316100 and the NSF of China under grants 61322102 and 61329101.}
\thanks{Copyright (c) 2014 IEEE. Personal use of this material is permitted.  However, permission to use this material for any other purposes must be obtained from the IEEE by sending a request to pubs-permissions@ieee.org}
}
\maketitle

\begin{abstract}
This paper studies the achievable degrees of freedom (DoF) for multi-user multiple-input multiple-output (MIMO) two-way relay channels, where there are $K$ source nodes, each equipped with $M$ antennas, one relay node, equipped with $N$ antennas, and each source node exchanges independent messages with an \textit{arbitrary} set of other source nodes via the relay. By allowing an arbitrary information exchange pattern, the considered channel model is a unified one. It includes several existing channel models as special cases: $K$-user MIMO Y channel, multi-pair MIMO two-way relay channel, generalized MIMO two-way X relay channel, and $L$-cluster MIMO multiway relay channel. Previous studies mainly considered the achievability of the DoF cut-set bound $2N$ at the antenna configuration $N < 2M$ by applying signal alignment for network coding. This work aims to investigate the achievability of the DoF cut-set bound $KM$ for the case $N\geq 2M$. To this end, we first derive tighter DoF upper bounds for three special cases of the considered channel model. Then, we propose a new transmission framework, \textit{generalized signal alignment} (GSA), to approach these bounds. The notion of GSA is to form network-coded symbols by aligning every pair of signals to be exchanged in a \textit{compressed} subspace at the relay. A necessary and sufficient condition to construct the relay compression matrix is given. We show that using GSA, the new DoF upper bound is achievable when i) $\frac{N}{M} \in \big(0, 2+\frac{4}{K(K-1)}\big] \cup \big[K-2, +\infty\big)$ for the $K$-user MIMO Y channel; ii) $\frac{N}{M} \in \big(0, 2+\frac{4}{K}\big] \cup \big[K-2, +\infty\big)$ for the multi-pair MIMO two-way relay channel; iii) $\frac{N}{M} \in \big(0, 2+\frac{8}{K^2}\big] \cup \big[K-2, +\infty\big)$ for the generalized MIMO two-way X relay channel. We also provide the antenna configuration regions for the general multi-user MIMO two-way relay channel to achieve the total DoF $KM$.
\end{abstract}


\section{Introduction}
Wireless relay has been an important ingredient in both \textit{ad hoc} and infrastructure-based wireless networks \cite{Gastpar,Kramer}. It shows great promises in power reduction, coverage extension and throughput enhancement. In the simplest scenario, a relay only serves a single user. This forms the classic relay channel, which includes one source, one destination and one relay. Nowadays, a relay has become very much like a wireless gateway where multiple users share a common relay and communicate with each other. A typical representative is the two-way relay channel (TWRC) \cite{Rankov1,Rankov,Wang2}, where two users exchange information with each other through a relay. A fundamental question that arises is what is the maximum number of data streams the relay can forward and how to achieve it. This motivates the analysis of degrees of freedom (DoF) and also drives the development of advanced relay strategies for efficient multi-user information exchange in the literature.

The success of the two-way relay channel owes to the invention of physical layer network coding (PLNC) \cite{ZhangSL,Katti}, which can almost double the spectral efficiency compared with traditional one-way relaying \cite{Wilson,Vaze,Yuan2}. In specific, when each source node is equipped with $M$ antennas and the relay node is equipped with $N$ antennas, the maximum achievable DoF of the multiple-input multiple-output (MIMO) two-way relay channel is $2\min\{M,N\}$ \cite{Vaze}. When three or more users arbitrarily exchange information with each other via a common relay, it is difficult to design PLNC due to multi-user interference and hence the analysis of DoF becomes challenging. Several multi-user MIMO two-way relay channels have been investigated in the literature, such as the MIMO Y channel \cite{Lee1,Chaaban1}, $K$-user MIMO Y channel \cite{Lee}, multi-pair MIMO two-way relay channel \cite{Ganesan,Chen,Avestimehr,Sezgin1}, MIMO two-way X relay channel \cite{Xiang}, generalized MIMO two-way X relay channel \cite{Liu}, $L$-cluster $K$-user MIMO multiway relay channel \cite{Tian,Yuan} and etc.

Based on the idea of interference alignment \cite{Jafar,Maddah}, \textit{signal alignment} (SA) is first proposed in \cite{Lee1} to analyze the maximum achievable DoF for the MIMO Y channel, where three users exchange independent messages with each other via the relay. By jointly designing the precoders at each source node, SA\footnote{Throughout this paper, SA refers to the method proposed in \cite{Lee1}.} is able to align the signals from two different source nodes in a same subspace of the relay node. By doing so, the two data streams to be exchanged between a pair of source codes are combined into one network-coded symbol and thus the relay can forward more data streams simultaneously. It is proved that with SA and network-coding aware interference nulling, the theoretical upper bound $3M$ of DoF is achievable when $N \geq \lceil \frac{3M}{2}\rceil$ \cite{Lee1}. Here, again, $M$ and $N$ denote the number of antennas at each source node and the relay node, respectively. The extension to $K$-user MIMO Y channels is considered in \cite{Lee}, where it is shown that the DoF upper bound is $\min\{KM,2N\}$ and the upper bound $2N$ in the case $N<2M$ is achievable when $N \leq \lfloor \frac{2K(K-1)M}{K(K-1)+2}\rfloor$. Here $K$ is the total number of users. The authors in \cite{Mu1} considered the case $N \geq 2M$ and showed that the upper bound $KM$ of DoF is achievable when $N \geq (K-1)M$ ($K$ is even) or $N \geq (K-1)M-1$ ($K$ is odd). The authors in \cite{Wang3} improved that result and showed that the upper bound $KM$ of DoF is achievable when $N \geq \lceil \frac{(K^2-2K)M}{K-1}\rceil$. Recently, the authors in \cite{Tian1} analyzed the multi-pair MIMO two-way relay channel and showed that the DoF upper bound $2N$ is achievable when $N \leq \lfloor\frac{2KM}{K+2}\rfloor$ and the DoF upper bound $KM$ is achievable when $N \geq KM$ \cite{Tian}. In \cite{Xiang}, SA is applied in the MIMO two-way X relay channel, where there are two groups of source nodes and one relay node, and each of the two source nodes in one group exchange independent messages with the two source nodes in the other group via the relay node. It is shown that the DoF upper bound is $2\min\{2M,N\}$, and the upper bound $2N$ is achievable when $N \leq \lfloor \frac{8M}{5} \rfloor$ by applying SA and interference cancellation. Despite the extensive work on this topic, the DoF achievebility of multi-user MIMO two-way relay channels still remains open in general.

In this paper, we are interested in the analysis of the DoF upper bound and the achievable DoF of a multi-user MIMO two-way relay channel for the antenna configuration $N \geq 2M$. In our considered multi-user MIMO two-way relay channel, there are $K$ source nodes each equipped with $M$ antennas, one relay node equipped with $N$ antennas, and each source node can arbitrarily select one or more partners to conduct independent information exchange. By allowing arbitrary information exchange pattern, our considered multi-user MIMO two-way relay channel is a unified channel model\footnote{We consider only ``unicast'' message exchange , i.e., the information to be exchanged is only limited within two users. Thus, we use the term ``two-way'' in our channel model, same as in \cite{ZhangSL,Ganesan,Xiang}. The authors in \cite{Tian} used the word ``multi-way'' to represent the same unicast message exchange between users, and the authors in \cite{Wang3} described it more explicitly as ``multiway with pairwise data exchange''. On the other hand, the ``multiway'' \cite{Gunduz} or ``multiway with clustered full data exchange'' \cite{Yuan} stand for ``multicast'' message, i.e. a common message is to be shared among more than two users.}. It includes several existing channel models as special cases, namely, $K$-user MIMO Y channel, multi-pair MIMO two-way relay channel, generalized MIMO two-way X relay channel and $L$-cluster $K{'}=\frac{K}{L}$-user MIMO multiway channel.

It is worth mentioning that SA is no longer feasible under the antenna configuration $N \geq 2M$. The reason is shown as follows. Recall that the SA condition \cite{Lee1} is
\begin{equation}\label{SA}
{{\bf H}}_{1,r}{{\bf V}}_{1}={{\bf H}}_{2,r}{{\bf V}}_{2},
\end{equation}
where ${\textbf H}_{i,r}$ is an $N \times M$ channel matrix from source $i$ to relay and ${\textbf V}_i$ is an $M \times d_{i,3-i}$ beamforming matrix of source $i$, for $i=1,2$, where $d_{i,3-i}$ denotes the number of data streams transmitted from source node $i$ to source node $3-i$. The above alignment condition can be rewritten as

\begin{equation}\label{SA_1}
\left[{{\bf H}}_{1,r}~~ -{{\bf H}}_{2,r} \right]
\left[
\begin{array}{ccc}
{{\bf V}}_{1} \\
{{\bf V}}_{2} \\
\end{array}
\right]
=0.
\end{equation}
Clearly, for the above equality to hold, one must have $N<2M$.

To achieve the maximum DoF at $N \geq 2M$ for multi-user MIMO two-way relay channels, it is not always optimal for users to utilize all the antennas at the relay. Specifically, using only a subset of antennas at the relay, known as \textit{antenna deactivation} \cite{Wang3}, can achieve higher DoF  for some cases \cite{Lee1,Tian}. But there is still a gap to the DoF cut-set bound. In this work, we first derive a tighter DoF upper bound and then we propose a new transmission framework, named \textit{generalized signal alignment} (GSA), which can achieve the DoF upper bound even when $N \geq 2M$. Compared with the conventional SA \cite{Lee1}, the proposed GSA has the following major difference. The signals to be exchanged do not align directly in the subspace observed by the relay. Instead, they are aligned in a compressed subspace after certain processing at the relay. This is done by jointly designing the precoding matrices at the source nodes and the compression matrix at the relay node. Compared with the existing alignment schemes \cite{Mu1,Wang3}, where the transmit precoding matrices and the relay processing matrix were also designed jointly, our proposed GSA differs fundamentally in the design methodology. In specific, the previous work first designed the transmit precoding matrices at each source node so that the received signal at the relay can form a pre-specified pattern, and then designed the relay processing matrix so that the network-coded symbols can be obtained from that pattern. As the pattern is pre-specified, the maximum achievable DoF by those previous schemes is also limited. On the other hand, we first design the processing matrix at the relay and then design each transmit precoding matrices. As a result, the signal received at the relay does not need to have any pattern. This leads to a higher achievable DoF than the previous results and makes the DoF upper bound tight at more antenna configurations of $\frac{N}{M}$.

The main contributions and results obtained in this work can be summarized as follows:
\begin{itemize}
  \item New DoF upper bounds are derived via genie-aided message approach for three special cases of multi-user MIMO two-way relay channel models, including the $K$-user MIMO Y channel, the multi-pair MIMO two-way relay channel, and the generalized MIMO two-way X relay channel. They are tighter than the cut-set bound.
  \item A new transmission framework, generalized signal alignment, is proposed. Its main idea is to align every pair of signals to be exchanged at a compressed subspace of the relay. A necessary and sufficient condition to construct the relay compression matrix is given. The proposed GSA represents a new and effective approach of integrating interference alignment with physical layer network coding towards the DoF analysis.
  \item The total DoF of $\sum_{i=1}^{K} \sum_{j \in {\cal S}_i} {d_{i,j}}$ is achievable when $M \geq \max_{i}\{\sum_{j \in {\cal S}_i} {d_{i,j}}\}$ and $N\geq (K-2)M+\max\{d_{i,j}\}$, where ${\cal S}_i$ is the set of source nodes that source node $i$ wishes to exchange messages with, and
       $d_{i,j}$ is the number of data streams to be transmitted from source node $i$ to source node $j$ for $j\in {\cal S}_i$.
        In particular, the total DoF upper bound $KM$ is achievable when $ \sum_{j \in {\cal S}_i} {d_{i,j}}=M$ for all $i$'s, and $N\geq (K-2)M+\max\{d_{i,j}\}$.
  \item For the special case of the $K$-user MIMO Y channel, by using GSA, the new DoF upper bound is tight when $\frac{N}{M} \in \big(0, 2+\frac{4}{K(K-1)}\big] \cup \big[K-2, +\infty\big)$.
  \item For the special case of the multi-pair MIMO two-way relay channel, by using GSA, the new DoF upper bound is tight when $\frac{N}{M} \in \big(0, 2+\frac{4}{K}\big] \cup \big[K-2, +\infty\big)$.
  \item For the special case of the generalized MIMO two-way X relay channel, by using GSA, the new DoF upper bound is tight when $\frac{N}{M} \in \big(0, 2+\frac{8}{K^2}\big] \cup \big[K-2, +\infty\big)$.
  \item For the special case of the $L$-cluster $K{'}$-user MIMO multiway relay channel, by using GSA, the DoF cut-set bound $KM$ is tight when $\frac{N}{M} \geq \frac{(K{'}-1)(K-2)+1}{K{'}-1}$.
\end{itemize}

The remainder of the paper is organized as follows. In Section II, we introduce the multi-user MIMO two-way relay channel. In Section III, we derive the tighter DoF upper bounds for $K$-user MIMO Y channel, multi-pair MIMO two-way relay channel, and generalized MIMO two-way X relay channel. In Section IV, we first introduce the principle of GSA transmission scheme, then give an illustrative example for the 4-user MIMO Y channel and finally compare our GSA with the existing transmission schemes. In Section V, we first apply the GSA transmission scheme to $K$-user MIMO Y channel, multi-pair MIMO two-way relay channel, and generalized MIMO two-way X relay channel to analyze their achievable DoF, and then we apply it to the general multi-user MIMO two-way relay channels. Section VI presents concluding remarks.

Notations: $(\cdot)^{T}$ and $(\cdot)^{H}$ denote the transpose and the Hermitian transpose, respectively. tr({\bf X}) and rank({\bf X}) stand for the trace and rank of {\bf X}. $\varepsilon[\cdot]$ stands for expectation. $\mbox{span} ({\bf X})$ and ${\mbox{null} ({\bf X})}$ stand for the column space and the null space of the matrix ${\bf X}$, respectively. $\mbox{dim}({\bf X})$ denotes the dimension of the column space of ${\bf X}$. $\lfloor x \rfloor$ denotes the largest integer no greater than $x$. $\lceil x \rceil$ denotes the smallest integer no less than $x$. {\bf I} is the identity matrix. $\left[{\bf X}\right]_{i,j}$ denotes the $(i,j)$-th entry of the matrix $\bf X$.

\section{System Model}
We consider a multi-user MIMO two-way relay channel, which consists of $K$ source nodes, each equipped with $M$ antennas, and one relay node, equipped with $N$ antennas. Each source node $i$, for $1 \leq i \leq K$, can exchange independent messages with an arbitrary set of other source nodes, denoted as ${\cal S}_i$, with the help of the relay. The message transmitted from source node $i$ to source node $j$, if $j \in {\cal S}_i$, is denoted as $W_{i,j}$ and it is independent for different $i$ and $j$. At each time slot, the message is encoded into a $d_{i,j} \times 1$ symbol vector $\textbf{s}_{i,j}=[s_{i,j}^1,s_{i,j}^2,\cdots,s_{i,j}^{d_{i,j}}]^T$, where $d_{i,j}$ denotes the number of independent data streams transmitted from source node $i$ to source node $j$. We define a $K \times K$ matrix ${\bf D}$, named as \textit{data switch matrix}, whose $(i,j)$-th entry is given by
\begin{equation}\label{D}
\left[{\bf D}\right]_{i,j}=\left\{
                \begin{array}{ll}
                  d_{i,j}, & \hbox{$j \in {\cal S}_i$, $\forall i$,} \\
                  0, & \hbox{otherwise.}
                \end{array}
              \right.
\end{equation}
Note that all the diagonal elements of $\bf D$ are zero. When the off-diagonal element $\left[{\bf D}\right]_{i,j}=0$, it means there is no information exchange between source node $i$ and $j$. In general, the data switch matrix $\bf D$ is not necessary to be symmetric. But for the convenience of analysis later, we only consider symmetric ${\bf D}$.
\begin{figure}[t]
\begin{centering}
\includegraphics[scale=0.4]{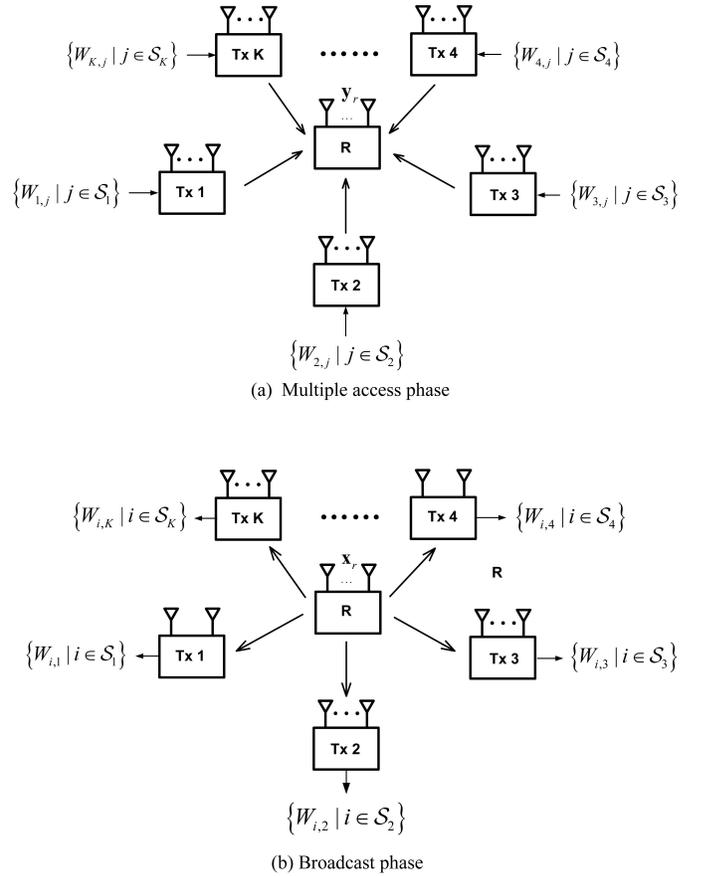}
\vspace{-0.1cm}
 \caption{Multi-user MIMO two-way relay channel}\label{Ge_MIMO_A_MAC_BC}
\end{centering}
\vspace{-0.2cm}
\end{figure}

The considered multi-user MIMO two-way relay channel is general in the sense that it includes the following existing channels as special cases:
\begin{itemize}
  \item The $K$-user MIMO Y channel: For each source node $i$, one has ${\cal S}_i=\{1,2, \cdots, K\}\backslash \{i\}$. The off-diagonal entries of $\bf D$, $\{d_{i,j} \mid i \neq j\}$, can be any nonnegative integer.
  \item The multi-pair MIMO two-way relay channel: Each source node $i$, $1 \leq i \leq \frac{K}{2}$, exchanges independent messages with its pair node $K+1-i$, and there are $\frac{K}{2}$ pairs in total. The entries of $\bf D$ which satisfy $\{\left[{\bf D}\right]_{i,j} \mid i+j \neq K+1\}$ must be zero. The rest can be any nonnegative integer.
  \item The generalized MIMO two-way X relay channel: The $K$ source nodes are divided into two groups. Each source node $i$ in one group exchanges independent messages with every source node in the other group. That is, ${\cal S}_i=\{j \mid \frac{K}{2}+1 \leq j \leq K\}$ for $1 \leq i \leq \frac{K}{2}$ and ${\cal S}_i=\{j \mid 1 \leq i \leq \frac{K}{2}\}$ for $\frac{K}{2}+1 \leq i \leq K$. The entries of $\bf D$ which satisfy $\{\left[{\bf D}\right]_{i,j} \mid 1 \leq i,j \leq \frac{K}{2}~\textrm{or}~\frac{K}{2}+1 \leq i,j \leq K\}$ must be zero. The rest can be any nonnegative integer.
\end{itemize}

The difference of these channels lies at the position of ``0'' in the data switch matrix $\bf D$. In this work, we unify these system models to the multi-user MIMO two-way relay channel.

The communication of the total messages takes place in two phases as shown in Fig. \ref{Ge_MIMO_A_MAC_BC}: the multiple access (MAC) phase and the broadcast (BC) phase. In the MAC phase, all $K$ source nodes transmit their signals to the relay simultaneously. Let ${\bf x}_i$ denote the transmitted signal vector from source node $i$. It is given by
\begin{eqnarray}\label{x_1}
{{\bf x}}_i =\sum\limits_{j \in {\cal S}_i} {\bf V}_{i,j}{\bf s}_{i,j}= {\bf V}_i{\textbf{s}}_i,
\end{eqnarray}
where ${\bf V}_{i,j}$ is the $M \times d_{i,j}$ precoding matrix for the information symbol vector ${\bf s}_{i,j}$ to be sent to source node $j$, ${\bf V}_i$ is a matrix obtained by stacking $\{{\bf V}_{i,j} \mid j \in {\cal S}_i\}$ by column and ${\bf s}_i$ is a vector obtained by stacking $\{{\bf s}_{i,j} \mid j \in {\cal S}_i\}$ by row. Each transmitted signal ${\bf x}_i$, for $i=1,~\cdots,~K$, satisfies the power constraint of
\begin{equation}\label{Power_constraint_source}
\textrm{tr}\left({\bf x}_i{\bf x}_i^H\right) \leq P,~\forall i
\end{equation}
where $P$ is the maximum transmission power allowed at each source node.

The received signal ${\bf y}_r$ at the relay is given by
\begin{eqnarray}\label{y_r}
{{\bf y}}_r=\sum\limits_{i=1}^{K}{{\bf H}}_{i,r}{{\bf x}}_{i}+{{\bf n}}_r,
\end{eqnarray}
where ${{\bf H}}_{i,r}$ denotes the frequency-flat quasi-static $N \times M$ complex-valued channel matrix from source node $i$ to the relay and ${{\bf n}}_r$ denotes the $N\times 1$ additive white Gaussian noise (AWGN) vector with each element being independent and having zero mean and unit variance.

In the BC phase, upon receiving ${{\bf y}}_r$ in \eqref{y_r}, the relay processes it to obtain a mixed signal ${\bf x}_r$, and broadcasts to all the users. The transmitted signal ${\bf x}_r$ satisfies the power constraint of
\begin{equation}\label{Power_constraint_relay}
\textrm{tr}\left({\bf x}_r{\bf x}_r^H\right) \leq P_r,
\end{equation}
where $P_r$ is the maximum transmission power allowed at the relay. Without loss of generality from the perspective of DoF analysis, we let $P_r=P$. The received signal at source node $i$ can be written as
\begin{eqnarray}\label{y_i}
{{\bf y}}_i={\bf G}_{r,i}{\bf x}_r+{{\bf n}}_i,
\end{eqnarray}
where ${{\bf G}}_{r,i}$ denotes the frequency-flat quasi-static $M \times N$ complex-valued channel matrix from relay to the source node $i$, and ${\bf n}_i$ denotes the AWGN at the source node $i$ with each element being independent and having zero mean and unit variance.  Each user tries to obtain its desired signal from its received signal using its own transmit signal as side information.

It is assumed that the channel state information $\{{\bf H}_{i,r}, {\bf G}_{r,i}\}$ is perfectly known at all source nodes and the relay, following the convention in \cite{Lee1,Lee,Wang3,Xiang,Tian}. The entries of the channel matrices are independent and identically distributed (i.i.d.) zero-mean complex Gaussian random variables with unit variance. Thus, each channel matrix has full rank with probability $1$. All the nodes in the network are assumed to be full duplex.

\section{New DoF Upper Bounds}
In this section, we first review the definition of DoF and the cut-set bound of DoF for the general multi-user MIMO two-way relay channel. After that we derive new DoF upper bounds for a set of special cases of channel models, namely, $K$-user MIMO Y channel, multi-pair MIMO two-way relay channel, and generalized MIMO two-way X relay channel.

Let $R_{i,j}$ denote the information rate carried in $W_{i,j}$. Since we assume the noise is i.i.d. zero-mean complex Gaussian random variables with unit variance, the average received signal-to-noise ratio (SNR) of each link is $P$. We define the DoF of the transmission from source node $i$ to source node $j$, for $j \in {\cal S}_i$, as
\begin{equation}\label{dof_source}
d_{i,j} \triangleq \lim\limits_{\textrm{SNR} \rightarrow \infty} \frac{R_{i,j}(\textrm{SNR})}{\textrm{log}(\textrm{SNR})}=\lim\limits_{P \rightarrow \infty} \frac{R_{i,j}(P)}{\textrm{log}(P)}.
\end{equation}

The DoF definition in \eqref{dof_source} captures the number of independent data streams transmitted from source node $i$ to source node $j$ and hence is the same as $d_{i,j}$ defined in the previous section. Then the total DoF of the system is
\begin{equation}\label{dof_all}
d_{total}=\sum\limits_{i=1}^K \sum\limits_{j \in {\cal S}_i} d_{i,j}.
\end{equation}

By applying the cut-set theorem \cite{Cover}, the total DoF upper bound of the multi-user MIMO two-way relay channel is given in the following lemma.

\textit{Lemma 1}: The total DoF of the multi-user MIMO two-way relay channel is upper-bounded by $\min \{KM,2N\}$.

\begin{proof}
The DoF upper bound of the source node $i$ is
\begin{align}\nonumber
\sum\limits_{j \in {\cal S}_i}^K d_{i,j} &\leq d_i^{upper} \\\nonumber
&= \min\left\{\underbrace{\min\{M,N\},}_{A}~\underbrace{\min\{(K-1)M,N\}}_{B}\right\}\\\label{d_per}
&=\min\{M, N\},
\end{align}
where $A$ is the bound for the cut from source node $i$ to the relay and $B$ is the bound for the cut from the relay to all the other source nodes.
Then
\begin{equation}\label{d_total_1}
d_{total} \leq \sum\limits_{i=1}^K d_i^{upper}=\min\{KM, KN\}.
\end{equation}

On the other hand, we consider the cut, denoted as $C$, from all the $K$ source nodes to the relay node at the MAC phase and the cut, denoted as $D$, from the relay node to all the $K$ source nodes at BC phase. We can obtain that
\begin{equation}\label{d_total_2}
d_{total} \leq \underbrace{\min\{KM,N\}}_{C}+\underbrace{\min\{KM,N\}}_{D}=\min\{2KM, 2N\}.
\end{equation}

Combining \eqref{d_total_1} and \eqref{d_total_2}, we obtain that
\begin{equation}\label{d_total}
d_{total} \leq \min\{KM, KN, 2KM, 2N\}=\min \{KM,2N\}.
\end{equation}
\end{proof}

Next, we present tighter DoF upper bounds for three special channel models by using the similar genie-aided approach in \cite{Wang3,Wang4}.

\textit{Theorem 1}: The total DoF for the $K$-user MIMO Y channel is piece-wise upper-bounded by \eqref{dof_KY_upper_information},  
\begin{figure*}[ht]
\begin{align}\label{dof_KY_upper_information}
d_{total} \leq \left\{
                    \begin{array}{ll}
                      2N, & \hbox{$\frac{N}{M} \in \left(0, \frac{2K^2-2K}{K^2-K+2}\right]$,} \\
                      \frac{2\beta K(K-1)M}{K(K-1)+\beta (\beta-1)}, & \hbox{$\frac{N}{M} \in \left(\frac{\beta(K(K-1)+(\beta-1)(\beta-2))}{K(K-1)+\beta (\beta-1)}, \beta\right]$,} \\
                      \frac{2K(K-1)N}{K(K-1)+\beta (\beta-1)}, & \hbox{$\frac{N}{M} \in \left(\beta, \frac{(\beta+1)(K(K-1)+\beta(\beta-1))}{K(K-1)+(\beta+1)\beta}\right]$,} \\
                      KM, & \hbox{$\frac{N}{M} \in \left(\frac{K^2-3K+3}{K-1}, +\infty \right)$.}
                    \end{array}
                  \right.
\end{align}
\hrule
\end{figure*}
where $\beta \in \{2,3,4,\cdots,K-2\}$.
\begin{proof}
See Appendix A.
\end{proof}

\textit{Theorem 2}: The total DoF for the multi-pair MIMO two-way relay channel is piece-wise upper-bounded by
\begin{align}\label{dof_MP_upper_information}
d_{total} \leq \left\{
                    \begin{array}{ll}
                      2N, & \hbox{$\frac{N}{M} \in \left(0, \frac{2K}{K+2}\right]$,} \\
                      \frac{2\beta KM}{K+\beta}, & \hbox{$\frac{N}{M} \in \left(\frac{\beta(K+\beta-2)}{K+\beta}, \beta\right]$,} \\
                      \frac{2KN}{K+\beta}, & \hbox{$\frac{N}{M} \in \left(\beta, \frac{(\beta+2)(K+\beta)}{K+\beta+2}\right]$,} \\
                      KM, & \hbox{$\frac{N}{M} \in \left(K-1, +\infty \right)$,}
                    \end{array}
                  \right.
\end{align}
where $\beta$ is an even number and $\beta \in \{2,4,\cdots,K-2\}$.
\begin{proof}
See Appendix B.
\end{proof}

\textit{Theorem 3}: The total DoF for the generalized MIMO two-way X relay channel is piece-wise upper-bounded by
\begin{align}\label{dof_X_upper_information}
d_{total} \leq \left\{
                    \begin{array}{ll}
                      2N, & \hbox{$\frac{N}{M} \in \left(0, \frac{2K^2}{K^2+4}\right]$,} \\
                      \frac{2K^2\beta M}{K^2+\beta ^2}, & \hbox{$\frac{N}{M} \in \left(\frac{(K^2+(\beta-2)^2)\beta}{K^2+\beta^2}, \beta\right]$,} \\
                      \frac{2K^2N}{K^2+\beta ^2}, & \hbox{$\frac{N}{M} \in \left(\beta, \frac{(K^2+\beta^2)(\beta+2)}{K^2+(\beta+2)^2}\right]$,} \\
                      KM, & \hbox{$\frac{N}{M} \in \left(\frac{K^2-2K+2}{K}, +\infty \right)$,}
                    \end{array}
                  \right.
\end{align}
where $\beta$ is an even number and $\beta \in \{2,4,\cdots,K-2\}$.
\begin{proof}
See Appendix C.
\end{proof}

The above new bounds will be shown to be tight at certain antenna configurations in Section V.

\section{Generalized Signal Alignment}
As mentioned in the introduction, the conventional SA in \cite{Lee1} is not feasible at the antenna configuration $N \geq 2M$. Thus, more advanced transmission strategies are desired. Recently, Wang and Yuan in \cite{Wang3}, and Mu and Tugnait in \cite{Mu1} proposed two different transmission frameworks, signal pattern and signal group based alignment, to analyze the achievable DoF when $N \geq 2M$ for $K$-user MIMO Y channel and MIMO multiway relay channel, respectively. However, there is still a gap between the achievable DoF and the best-known upper bound. In this work, we propose a new transmission framework, named as \textit{generalized signal alignment}, based on which we will study the DoF achievability of the general multi-user MIMO two-way relay channel. In this section, we first introduce the basic principle of GSA and then give an example. After that we present the difference with existing schemes.

\subsection{Basic principle}
We rewrite the received signal \eqref{y_r} at the relay during the MAC phase as
\begin{align}\label{y_r_align}
{{\bf y}}_r=\sum\limits_{i=1}^K {\bf H}_{i,r}{\bf V}_i\textbf{s}_i+{\bf n}_r.
\end{align}
Note that the total number of independent data streams to communicate is $d_{total}=\sum_{i=1}^{K} \sum_{j \in {\cal S}_i} {d_{i,j}}$. When $N \geq d_{total}$, the relay can decode all the data streams and the decode-and-forward (DF) relay is the optimal strategy. When $N < d_{total}$, it is impossible for the relay to decode all the data streams individually. However, applying physical layer network coding, we only need to obtain the network-coded symbol vector at the relay, denoted as $\textbf{s}_{\oplus}$, where $\textbf{s}_{\oplus}$ is a vector obtained by stacking the $\{{\bf s}_{i,j}+{\bf s}_{j,i},~\forall j \in {\cal S}_i,~\forall i\}$ by row.

According to the signal alignment equation \eqref{SA}, when $N \geq 2M$, $\textbf{s}_{\oplus}$ cannot be obtained directly by designing the precoding matrices ${\bf V}_{i,j}$ and ${\bf V}_{j,i}$. Instead of aligning the signals to be exchanged directly at a same subspace of the relay, we propose to align them at a same compressed subspace of the relay. This is realized by the joint design of the source precoding matrices and relay processing matrix. Mathematically, let ${\bf P}$ denote a $J \times N$ ($J \leq N$) full-rank compression matrix, then the received signal at the relay after compression can be written as
\begin{align}\label{y_r_projection}
\hat{{\bf y}}_r={{\bf P}}{{\bf y}}_r=\sum\limits_{i=1}^K {\bf P}{\bf H}_{i,r}{\bf V}_i\textbf{s}_i+{\bf P}{\bf n}_r,
\end{align}
The proposed \textit{generalized signal alignment equation} is given by
\begin{align}\label{GSA_constraints}
{\bf P}{\bf H}_{i,r}{\bf V}_{i,j}={\bf P}{\bf H}_{j,r}{\bf V}_{j,i} \triangleq {\bf B}_{i,j}, ~\forall i,j~\textrm{with}~\left[{\bf D}\right]_{i,j} \neq 0.
\end{align}
Note that all ${\bf V}_{i,j}$ should have full rank in order to ensure the decodability of the network-coded symbol vector $\textbf{s}_{\oplus}$ at the relay. Moreover, the compression via $\bf P$ should not sacrifice the decodability of these messages. The GSA equation \eqref{GSA_constraints} can also be rewritten as
\begin{align}\label{GSA_2}
\left[{\bf P}{\bf H}_{i,r}~-{\bf P}{\bf H}_{j,r}\right]
\left[ \begin{array}{c}
        {\bf V}_{i,j} \\
        {\bf V}_{j,i}
      \end{array}
\right]&=0
\end{align}
or equivalently
\begin{align}\label{GSA_v_a}
\left[\begin{array}{c}
        {\bf V}_{i,j} \\
        {\bf V}_{j,i}
      \end{array}
\right] \subseteq \textbf{Null}~\left[{\bf P}{\bf H}_{i,r}~-{\bf P}{\bf H}_{j,r}\right].
\end{align}

If we can align each pair of data streams to be exchanged in a same subspace, the dimension of the received signal after compression $\hat{{\bf y}}_r$ must be no less than $\frac{d_{total}}{2}$ in order to guarantee the decodability of $\textbf{s}_{\oplus}$ at the relay. This indicates that $J \geq \frac{d_{total}}{2}$. In this paper, we assume that $J = \frac{d_{total}}{2}$. The compressed signal $\hat{{\bf y}}_r$ can be rewritten as
\begin{align}\label{y_r_projection}
\hat{{\bf y}}_r={\bf B}{\bf s}_{\oplus}+{\bf P}{\bf n}_r,
\end{align}
where $\bf B$ is a matrix obtained by stacking the ${\bf B}_{i,j}$ with $\left[{\bf D}\right]_{i,j} \neq 0$ by column.

\textit{Remark 1}: If the rank of $\bf P$ is less than $\frac{d_{total}}{2}$, then the $\frac{d_{total}}{2}$ number of network-coded symbols cannot be fully decodable. Hence, we assume that $\bf P$ is a full-rank matrix.

\textit{Remark 2}: Since the entries of all channel matrices are independent and Gaussian, the probability that a basis vector in the intersection space spanned by the effective channel matrices of one pair of source nodes (i.e. ${\bf PH}_{i,r}$ and ${\bf PH}_{j,r}$) lies in the intersection space of another pair is zero \cite{Lee1}. Thus, $\bf B$ is a full rank matrix with probability 1, which guarantees the decodability of $\textbf{s}_{\oplus}$ at the relay.

\textit{Remark 3}: Once we can obtain the network-coded symbol vector $\textbf{s}_{\oplus}$ at the relay during the MAC phase, each user can obtain its desired signals during the BC phase due to the symmetry between MAC phase and BC phase \cite{Jindal}.

\begin{figure*}[t]
\begin{centering}
\includegraphics[scale=0.5]{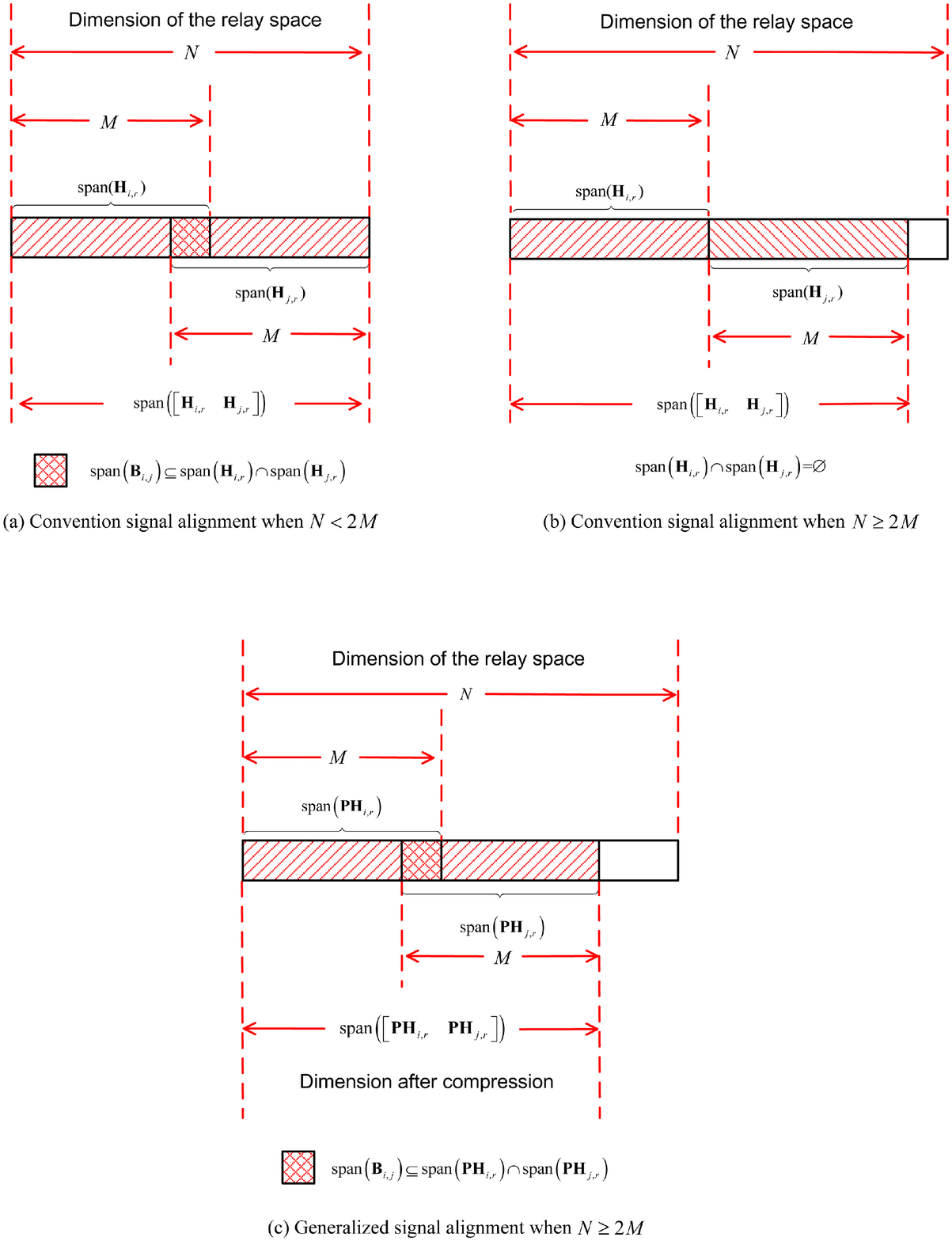}
\vspace{-0.1cm}
 \caption{Span space of ${\bf P}{\bf H}_{i,r}$ and ${\bf P}{\bf H}_{j,r}$.}\label{Fig_GSA_span}
\end{centering}
\vspace{0.2cm}
\hrule
\end{figure*}

Before discussing the condition when the GSA equation holds, we first analyze it from the span space perspective. From \eqref{GSA_constraints}, we can obtain that
\begin{align}\label{B_span}
&\textrm{span}\left({\bf B}_{i,j}\right) \subseteq \textrm{span} \left({\bf P}{\bf H}_{i,r}\right)\cap\textrm{span}\left({\bf P}{\bf H}_{j,r}\right).
\end{align}
Fig. \ref{Fig_GSA_span} provides an illustration of \eqref{B_span} through comparison with the conventional SA. It is seen that when $N \geq 2M$,  the intersection space between $\textrm{span}\left({\bf H}_{i,r}\right)$ and $\textrm{span}\left({\bf H}_{j,r}\right)$ is null if without compression. Only after compression, the intersection space will be non-empty and then signal alignment becomes possible.

In what follows, we provide the necessary and sufficient condition on the compression matrix $\bf P$ for the GSA equation to hold.

\textit{Theorem 4}: The GSA equation \eqref{GSA_constraints} holds if and only if there are at least $\frac{d_{total}}{2}-2M+d_{i,j}$ basis vectors of $\textrm{span} \left({\bf P}^T\right)$ that lie in the null space of $\left[{\bf H}_{i,r}~-{\bf H}_{j,r}\right]^T$ for any source pair $(i,j)$ with $\left[{\bf D}\right]_{i,j} \neq 0$.
\begin{proof}
First, we prove the \textit{only if} part. For any source pair $(i,j)$ with $\left[{\bf D}\right]_{i,j} \neq 0$, when the GSA equation \eqref{GSA_constraints}, or equivalently \eqref{GSA_v_a}, holds, then the dimension of the null space of $\left[{\bf P}{\bf H}_{i,r}~-{\bf P}{\bf H}_{j,r}\right]$ must be greater than or equal to $d_{i,j}$. That is
\begin{align}\nonumber
2M-\textrm{rank}\left(\left[{\bf P}{\bf H}_{i,r}~-{\bf P}{\bf H}_{j,r}\right]\right) \geq d_{i,j}
\end{align}
or equivalently
\begin{align}\label{rank_aij}
\textrm{rank}\left({\bf P}\left[{\bf H}_{i,r}~-{\bf H}_{j,r}\right]\right) &\leq 2M-d_{i,j}.
\end{align}
From \eqref{rank_aij}, it is seen that there must be an elementary matrix $\bf Q$ such that
\begin{equation}\label{span_PH}
{\bf Q}{\bf P}\left[{\bf H}_{i,r}~-{\bf H}_{j,r}\right]= \left[\begin{array}{cc}
                      {\bf 0}_{i,j}\\
                      {\bf \Lambda}_{i,j}
                    \end{array}
\right],
\end{equation}
where ${\bf 0}_{i,j}$ is a $(\frac{d_{total}}{2}-2M+d_{i,j}) \times N$ zero matrix and ${\bf \Lambda}_{i,j}$ is a $(2M-d_{i,j}) \times N$ matrix with rank at most $2M-d_{i,j}$. Since $\left[{\bf H}_{i,r}~-{\bf H}_{j,r}\right]$ is a full-rank matrix with probability 1 due to the property of random matrices, we assume that it always has full rank throughout this paper. Then, there must be at least $\frac{d_{total}}{2}-2M+d_{i,j}$ row vectors of ${\bf QP}$ that lie in the left null space of $\left[{\bf H}_{i,r}~-{\bf H}_{j,r}\right]$. Note that the basis vectors of $\textrm{span} \left({\bf P}^T\right)$ and $\textrm{span} \left({\bf P}^T{\bf Q}^T\right)$ are the same because $\bf Q$ is an elementary matrix. Hence, there are at least $\frac{d_{total}}{2}-2M+d_{i,j}$ basis vectors of $\textrm{span} \left({\bf P}^T\right)$ that lie in the null space of $\left[{\bf H}_{i,r}~-{\bf H}_{j,r}\right]^T$ for any source pair $(i,j)$ with $\left[{\bf D}\right]_{i,j} \neq 0$.

Then we prove the \textit{if} part. If there are at least $\frac{d_{total}}{2}-2M+d_{i,j}$ basis vectors of $\textrm{span} \left({\bf P}^T\right)$ that lie in the null space of $\left[{\bf H}_{i,r}~-{\bf H}_{j,r}\right]^T$ for any source pair $(i,j)$ with $\left[{\bf D}\right]_{i,j} \neq 0$, then we can construct at least $\frac{d_{total}}{2}-2M+d_{i,j}$ row vectors of $\bf P$ falling in the left null space of $\left[{\bf H}_{i,r}~-{\bf H}_{j,r}\right]$, resulting $\textrm{rank}({\bf P}\left[{\bf H}_{i,r}~-{\bf H}_{j,r}\right]) \leq 2M-d_{i,j}$. As a result, the dimension of the null space of $\left[{\bf P}{\bf H}_{i,r}~-{\bf P}{\bf H}_{j,r}\right]$ is no less than $d_{i,j}$. Thus, it is always feasible to find the precoding matrices ${\bf V}_{i,j}$ and ${\bf V}_{j,i}$ based on \eqref{GSA_v_a}. Hence, the GSA equation holds.
\end{proof}

\textit{Theorem 4} not only shows the necessary and sufficient condition on the relay compression matrix $\bf P$ for GSA, but also provides an insight into the joint design of the relay compression matrix $\bf P$ and the source precoding matrices ${\bf V}_{i,j}$. More specifically, from the proof of the \textit{if} part, it is seen that we should first construct the compression matrix $\bf P$ such that there are at least $\frac{d_{total}}{2}-2M+d_{i,j}$ row vectors of $\bf P$ that lie in the left null space of $\left[{\bf H}_{i,r}~-{\bf H}_{j,r}\right]$ for any source pair $(i,j)$ with $\left[{\bf D}\right]_{i,j} \neq 0$. After that, we should treat ${\bf PH}_{i,r}$ as the effective channel matrix from source node $i$ to the relay node and design the source precoding matrices ${\bf V}_{i,j}$ based on the conventional SA. As such, the main challenge in GSA is to construct $\bf P$.  In Section V, we will present a general guideline to design $\bf P$ and also present the specific construction of $\bf P$ for some special channel models.

\subsection{An example}
In this subsection, we use the $4$-user MIMO Y channel, a special case of the multi-user MIMO two-way relay channel, to demonstrate the GSA. We consider the simplest case with $M=3$ and $d_{i,j}=1$ for any $i \neq j$. The corresponding data switch matrix ${\bf D}$ is
\begin{equation}\label{D_Y_4}
{{\bf D}}=\left[\begin{array}{cccc}
                 0 & 1 & 1 & 1 \\
                 1 & 0 & 1 & 1 \\
                 1 & 1 & 0 & 1 \\
                 1 & 1 & 1 & 0
               \end{array}
\right].
\end{equation}
In what follows, we show how to implement GSA when there are $N=7$ antennas at the relay node.

We design a $6 \times 7$ compression matrix $\bf P$ at the relay as follows:
\begin{equation}\label{P_Y_4}
{\bf P}=\left[\begin{array}{c}
                {\bf p}_{1,2} \\
                {\bf p}_{1,3} \\
                {\bf p}_{1,4} \\
                {\bf p}_{2,3} \\
                {\bf p}_{2,4} \\
                {\bf p}_{3,4}
              \end{array}
\right]
\end{equation}
where
\begin{align}\nonumber
{\bf p}_{1,2}^T\subseteq \textbf{Null}~\big[{\bf H}_{3,r}~{\bf H}_{4,r}\big]^T,~~{\bf p}_{1,3}^T\subseteq \textbf{Null}~\big[{\bf H}_{2,r}~{\bf H}_{4,r}\big]^T,~\\\nonumber
{\bf p}_{1,4}^T\subseteq \textbf{Null}~\big[{\bf H}_{2,r}~{\bf H}_{3,r}\big]^T,~~{\bf p}_{2,3}^T\subseteq \textbf{Null}~\big[{\bf H}_{1,r}~{\bf H}_{4,r}\big]^T,~\\
{\bf p}_{2,4}^T\subseteq \textbf{Null}~\big[{\bf H}_{1,r}~{\bf H}_{3,r}\big]^T,~~{\bf p}_{3,4}^T\subseteq \textbf{Null}~\big[{\bf H}_{1,r}~{\bf H}_{2,r}\big]^T.~
\end{align}

Clearly, $\bf P$ satisfies \textit{Theorem 4}. Then, we design the precoding matrix ${\bf V}_{i,j}$ according to \eqref{GSA_v_a}. Thus, the GSA equation \eqref{GSA_constraints} holds.

\begin{figure}[t]
\begin{centering}
\includegraphics[scale=0.32]{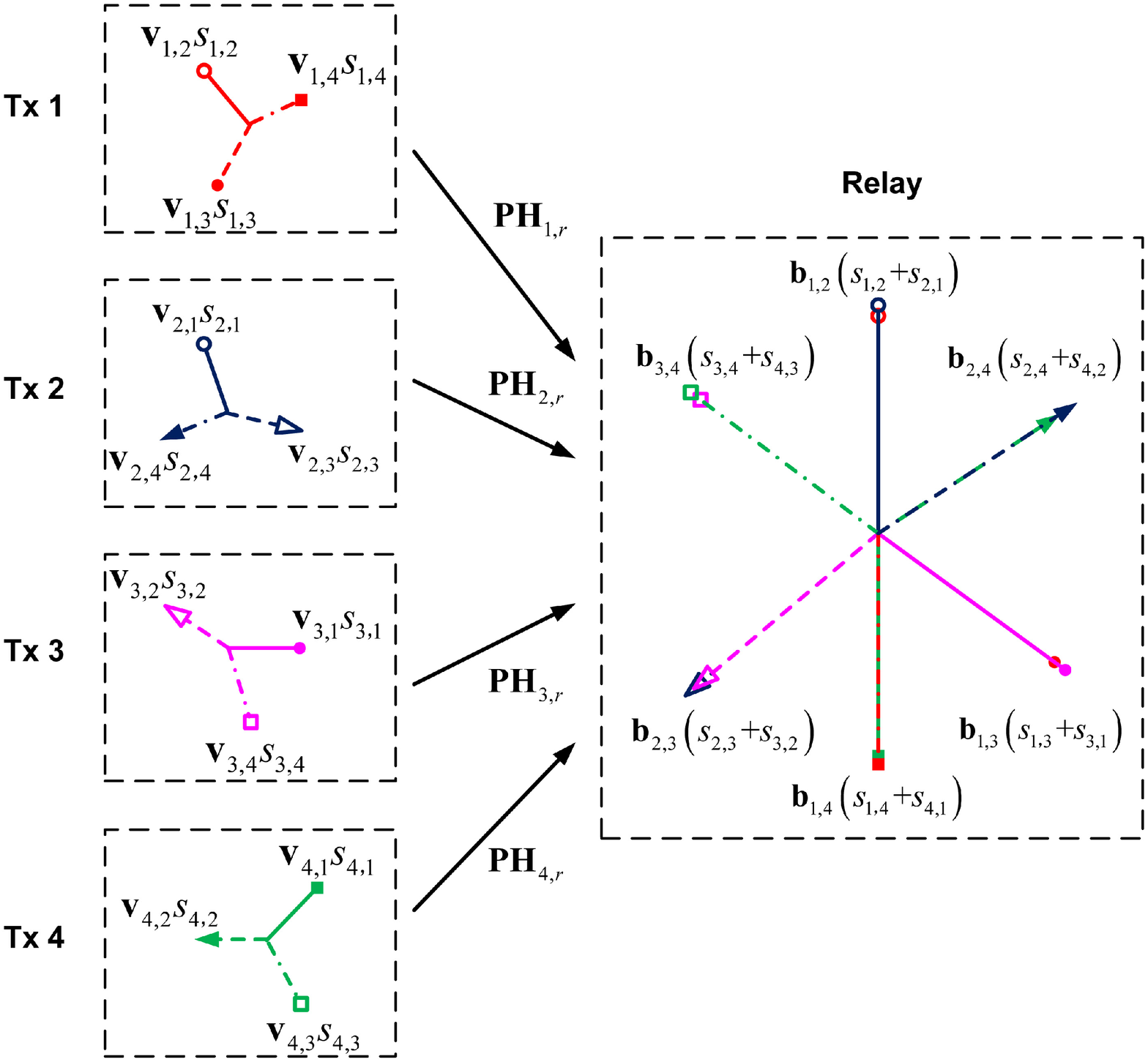}
\vspace{-0.1cm}
 \caption{Alignment in the MAC phase.}\label{GSA_V}
\end{centering}
\vspace{-0.2cm}
\end{figure}


After simple manipulation, the equivalent channel vector seen by $s_{1,2}$ and $s_{2,1}$ becomes
\begin{equation}\label{b_12_Y_4}
{\bf b}_{1,2}={\bf P}{\bf H}_{1,r}{\bf v}_{1,2}={\bf P}{\bf H}_{2,r}{\bf v}_{2,1}=\alpha_{1,2} [1~0~0~0~0~0]^T,
\end{equation}
where $\alpha_{1,2}$ is a constant.

Similarly, we can obtain all the other equivalent channel vectors as
\begin{align}\nonumber
{\bf b}_{1,3}={\bf P}{\bf H}_{1,r}{\bf v}_{1,3}={\bf P}{\bf H}_{3,r}{\bf v}_{3,1}=\alpha_{1,3} [0~1~0~0~0~0]^T,\\\nonumber
{\bf b}_{1,4}={\bf P}{\bf H}_{1,r}{\bf v}_{1,4}={\bf P}{\bf H}_{4,r}{\bf v}_{4,1}=\alpha_{1,4} [0~0~1~0~0~0]^T,\\\nonumber
{\bf b}_{2,3}={\bf P}{\bf H}_{2,r}{\bf v}_{2,3}={\bf P}{\bf H}_{3,r}{\bf v}_{3,2}=\alpha_{2,3} [0~0~0~1~0~0]^T,\\\nonumber
{\bf b}_{2,4}={\bf P}{\bf H}_{2,r}{\bf v}_{2,4}={\bf P}{\bf H}_{4,r}{\bf v}_{4,2}=\alpha_{2,4} [0~0~0~0~1~0]^T,\\\label{direction_Y_4}
{\bf b}_{3,4}={\bf P}{\bf H}_{3,r}{\bf v}_{3,4}={\bf P}{\bf H}_{4,r}{\bf v}_{4,3}=\alpha_{3,4} [0~0~0~0~0~1]^T.
\end{align}
Therefore, the overall received signals after compression at the relay can be written as
\begin{align}\label{y_r_projection_Y4}
\hat{{\bf y}}_r=\left[\begin{array}{c}
                        \alpha_{1,2}(s_{1,2}+s_{2,1}) \\
                        \alpha_{1,3}(s_{1,3}+s_{3,1}) \\
                        \alpha_{1,4}(s_{1,4}+s_{4,1}) \\
                        \alpha_{2,3}(s_{2,3}+s_{3,2}) \\
                        \alpha_{2,4}(s_{2,4}+s_{4,2}) \\
                        \alpha_{3,4}(s_{3,4}+s_{4,3})
                      \end{array}
\right]+{\bf P}{\bf n}_r.
\end{align}

Now we can see that the signal pairs $s_{i,j}$ and $s_{j,i}$ are not only aligned in the same dimension but also in orthogonal dimensions. Fig. \ref{GSA_V} illustrates the notion of GSA in the MAC phase where there are 6 network-coded symbols aligned at the relay. The following network-coded symbol vector can be readily estimated from $\hat{{\bf y}}_r$ at the relay:
\begin{eqnarray}\label{s_oplus}
\hat{{\bf s}}_{\oplus}=\left[\begin{array}{c}
                        s_{1,2}+s_{2,1} \\
                        s_{1,3}+s_{3,1} \\
                        s_{1,4}+s_{4,1} \\
                        s_{2,3}+s_{3,2} \\
                        s_{2,4}+s_{4,2} \\
                        s_{3,4}+s_{4,3}
                      \end{array}
\right].
\end{eqnarray}

\begin{figure}[t]
\begin{centering}
\includegraphics[scale=0.7]{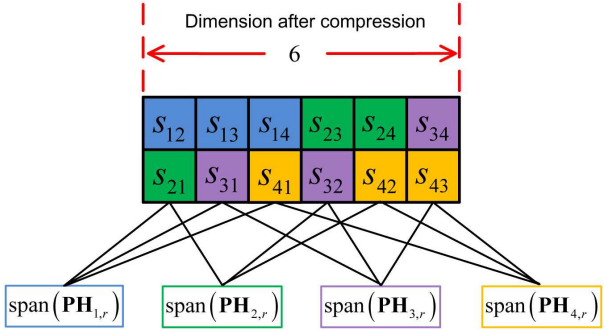}
\vspace{-0.1cm}
 \caption{Span space of each signal at the relay after compression.}\label{Fig_GSA_Y_4}
\end{centering}
\vspace{-0.2cm}
\end{figure}

Fig. \ref{Fig_GSA_Y_4} illustrates the space where each signal spans after compression. It can be seen that $\textrm{rank}(\left[{\bf P}{\bf H}_{i,r}~-{\bf P}{\bf H}_{j,r}\right])=5$ for any pair $(i,j)$ and the total dimension of the space at the relay after compression is 6.

\subsection{Comparison with the existing transmission schemes}
\begin{figure}[t]
\begin{centering}
\includegraphics[scale=0.45]{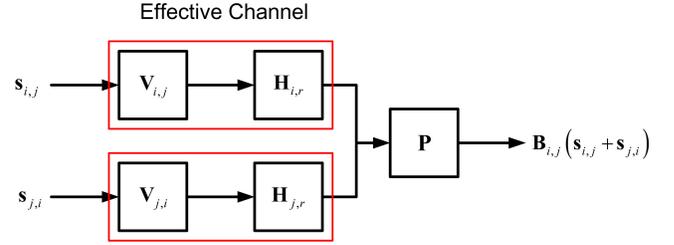}
\vspace{-0.1cm}
 \caption{Illustration of the schemes in \cite{Mu1} and \cite{Wang3}.}\label{Fig_signal_process_1}
\end{centering}
\vspace{-0.2cm}
\end{figure}

\begin{figure}[t]
\begin{centering}
\includegraphics[scale=0.45]{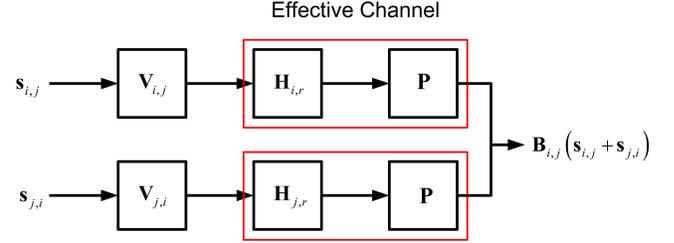}
\vspace{-0.1cm}
 \caption{Illustration of GSA.}\label{Fig_signal_process_2}
\end{centering}
\vspace{-0.2cm}
\end{figure}

Previously, there were two main transmission frameworks to analyze the DoF when $N \geq 2M$. The first method is proposed by Mu and Tugnait in \cite{Mu1}, named as \textit{signal group based alignment}. The second is proposed by Wang and Yuan in \cite{Wang3}, named as \textit{signal pattern}. The main idea of the two methods is to design the precoding matrices at each source node first under certain rules, such as group or pattern, and then to design the relay processing matrix so that the network-coded symbol vector $\textbf{s}_{\oplus}$ can be decoded from the signal received at the relay after processing. Here, when designing the relay processing matrix, the multiplication of each source precoding matrix and the channel matrix can be regarded as the effective channel matrix from each source to the relay. This is illustrated in Fig. \ref{Fig_signal_process_1}. Note that the designed $\bf P$ and ${\bf V}_{i,j}$ in both \cite{Mu1} and \cite{Wang3} also satisfy the GSA equation \eqref{GSA_constraints}.
In our proposed GSA, we first design the compression matrix at the relay, and then construct the precoding matrices at each source node by treating the compressed channel matrix as the effective channel matrix from each source node to the relay node. We illustrate the idea of the GSA in Fig. \ref{Fig_signal_process_2}.

By comparing Fig. \ref{Fig_signal_process_1} and Fig. \ref{Fig_signal_process_2}, it is seen that the main difference between the previous schemes and our GSA is that we reverse the design order of $\bf P$ at the relay node and $\{{\bf V}_{i,j}\}$ for each source node. Note that $\bf P$ is a common matrix for processing the signal at the relay and $\{{\bf V}_{i,j}\}$ is a set of private matrices designed for each source node. The existing schemes \cite{Mu1} and \cite{Wang3} first designed the transmit precoding matrices at each source node so that the received signal at the relay can form a pre-specified pattern and then designed the relay processing matrix so that the network-coded symbol can be obtained from that pattern. As the pattern is pre-specified, the maximum achievable DoF by those previous schemes is also limited. On the other hand, we first design the processing matrix at the relay and then design each transmit precoding matrices. As a result, the signal received at the relay does not need to have any pattern. This will lead to a higher achievable DoF than the previous results as shown in the next section.

\section{Analysis of DoF Achievability with Generalized Signal Alignment}
In this section, we first apply GSA in three special cases of multi-user MIMO two-way relay channels, including $K$-user MIMO Y channel, multi-pair MIMO two-way relay channel, and generalized MIMO two-way X relay channel. The DoF upper bound derived in Section III of each channel is proved to be tight under some specific regions of $\frac{N}{M}$. Then we apply GSA in the general case with arbitrary data switch matrix $\bf D$ and show that the DoF of $\sum\limits_{i=1}^{K} \sum\limits_{j \in {\cal S}_i} {d_{i,j}}$ is achievable when $N \geq (K-2)M+\max\{d_{i,j}\mid \forall i,j\}$. Finally, we extend the results to the $L$-cluster $K{'}=\frac{K}{L}$-user multiway relay channel.

To assist the DoF analysis, we first introduce a so-called \textit{DoF plane} to conveniently represent the DoF values at different antenna configurations and present a useful lemma. Then we will introduce a guideline to the construction of $\bf P$ in the GSA equation.

Assume that at antenna configuration $N=\alpha_0 M$, the total DoF $d_0M$ is achievable, where $\alpha_0$ and $d_0$ are some constants (which may depend on the number of source nodes, $K$). Then we say the point $Q=(\alpha_0, d_0)$ is achievable in a 2-dimensional DoF plane, where the $x$-axis is the ratio of the antenna configurations $\frac{N}{M}$, and the $y$-axis is the DoF value with respect to $M$. Alternatively, when a point  $Q=(\alpha_0, d_0)$ in the 2-dimensional DoF plane is achievable, it means the DoF $d_0M$ is achievable at the antenna configuration $N=\alpha_0 M$.

\begin{figure}[t]
\begin{centering}
\includegraphics[scale=0.38]{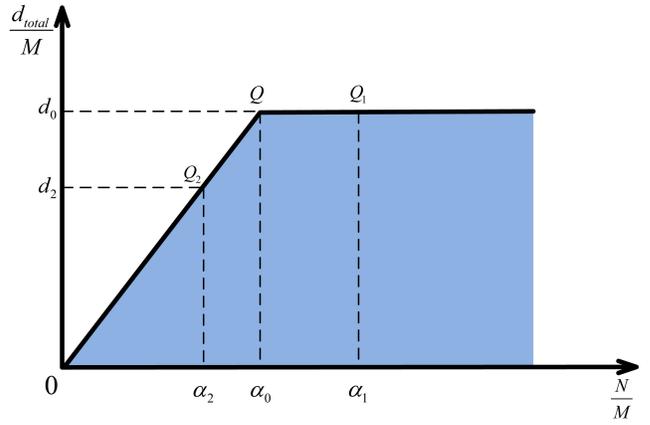}
\vspace{-0.1cm}
 \caption{Single-sided trapezoid characterized by $Q$ in the 2-dimensional DoF plane.}\label{lemma_2}
\end{centering}
\vspace{-0.2cm}
\end{figure}

\textit{Lemma 2}: If the point $Q=(\alpha_0, d_0)$ is achievable in the 2-dimensional DoF plane, then all points in the single-sided trapezoid characterized by $Q$, as shown in Fig. \ref{lemma_2}, are achievable.
\begin{proof}
To prove this lemma, we only need to show that $Q_1=(\alpha_1, d_0)$ and $Q_2=(\alpha_2, d_2)$ as plotted in Fig. \ref{lemma_2} are achievable. For $Q_1$, $N=\alpha_1 M>\alpha_0 M$, let the relay node only utilize $N{'}=\alpha_0 M$ antennas. Then the total DoF of $d_0 M$ should be achievable due to the achievability of $Q$. Thus, $Q_1$ is achievable. For $Q_2$, $N=\alpha_2 M < \alpha_0 M$ and $\frac{d_0}{\alpha_0}=\frac{d_2}{\alpha_2}$, let each source node only utilize $M{'}=\frac{N}{\alpha_0}$ antennas. Then the total DoF of $d_0 M{'}$ should be achievable, again, due to the achievability of $Q$. Since $d_0 M{'}=\frac{d_0 N}{\alpha_0}=\frac{d_2 N}{\alpha_2}=d_2 M$, this is equivalent to that $Q_2$ is achievable.

Note that if the number of antennas after deactivation is not an integer but a fraction $\frac{s}{t}$, then we can use the method of $t$-symbol extensions to achieve the total DoF.
\end{proof}

In what follows, we will show the guideline of the construction of $\bf P$. Define $\beta=\lfloor\frac{N}{M}\rfloor \geq 2$. First, select $\beta$ out of the $K$ source nodes and denote them as $\{\pi(1), \cdots, \pi(\beta)\}$. Then, define an $N \times \beta M$ $\beta$-combining channel matrix ${\cal H}_{\beta}$ \footnote{The $\beta$-combining matrix ${\cal H}_{\beta}$ is assumed to always have full rank throughout this paper since it is a full-rank matrix with probability 1 due to the property of random matrices.} such that
\begin{equation}
\textrm{span}\left({\cal H}_{\beta}\right)=\textrm{span} \left[{\bf H}_{\pi(1),r},{\bf H}_{\pi(2),r},\cdots,{\bf H}_{\pi(\beta),r}\right].
\end{equation}
In total, there are $\binom{K}{\beta}$ different $\beta$-combining channel matrices. We denote them as ${\cal H}_{\beta}[n]$, where $n=1,2,\cdots,\binom{K}{\beta}$. For each ${\cal H}_{\beta}[n]$, we choose $\kappa[n]$ out of the $N-\beta M$ basis vectors of the left null space of ${\cal H}_{\beta}[n]$ and treat them as the row candidates of $\bf P$. In total, we can have $\sum\limits_{i=1}^{\binom{K}{\beta}} \kappa[n]$ possible rows for $\bf P$. Note that the $\kappa[n]$ rows constructed from ${\cal H}_{\beta}[n]$ will be independent from the $\kappa[m]$ rows constructed from ${\cal H}_{\beta}[m]$. This can be proved by contradiction\footnote{Assume a vector $\tilde{{\bf p}}$ is located in the left null space of both ${\cal H}_{\beta}[n]$ and ${\cal H}_{\beta}[m]$, and ${\cal H}_{\beta}[n]$ and ${\cal H}_{\beta}[m]$ differ by at least one sub-matrix, say ${\bf H}_{i,r}$ as included ${\cal H}_{\beta}[m]$. Then $\tilde{{\bf p}}$ shall also be located in the left null space of $\left[{\cal H}_{\beta}[n], {\bf H}_{i,r}\right]$, which is an $N \times (\beta+1)M$ matrix. However, since $\beta M \leq N < (\beta+1) M$, the left null space of $\left[{\cal H}_{\beta}[n], {\bf H}_{i,r}\right]$ should be null. This contradicts the existence of such $\tilde{{\bf p}}$.}. It is also noted that for any source pair $(i,j)$ with $[{\bf D}]_{i,j} \neq 0$, as long as $\textrm{span}([{\bf H}_{i,r}, {\bf H}_{j,r}])$ is a subspace of $\textrm{span} ({\cal H}_{\beta}[n])$, then the $\kappa[n]$ basis vectors of the left null space of $\textrm{span} ({\cal H}_{\beta}[n])$ will also be orthogonal to $[{\bf H}_{i,r}, {\bf H}_{j,r}]$. So the next step in constructing $\bf P$ is to traverse all the $\textrm{span} ({\cal H}_{\beta}[n])$ that include $\textrm{span}([{\bf H}_{i,r}, {\bf H}_{j,r}])$ as a subspace and find at least $\frac{d_{total}}{2}-2M+d_{i,j}$ rows of $\bf P$ in order to meet the condition in \textit{Theorem 4}. This is realized by determining the value of each $\kappa[n]$. The specific realization method will be detailed in the proof of the DoF results in each of the following subsections.

\subsection{$K$-user MIMO Y channel}
The considered $K$-user MIMO Y channel consists of $K$ source nodes, each equipped with $M$ antennas, and one relay node, equipped with $N$ antennas. Each source node exchanges independent messages with all the other $K-1$ source nodes with the help of the relay.

\begin{table*}[tbp]
\tiny
\centering
\caption{Recent Advances towards the DoF Analysis for $K$-user MIMO Y Channel}\label{table1}
\begin{tabular}{|c|c|c|c|}
\hline
$K$ & $\frac{N}{M}$ & Maximum DoF & Reference\\ \hline
3 &  $\left[\frac{3}{2}, +\infty\right)$ & $3M$ & \cite{Lee1}\\ \hline
3 &  $\left(0,+\infty\right)$ & $\min\{3M,2N\}$ & \cite{Chaaban1} \\\hline
4 &  $\left(0,\frac{12}{7}\right] \cup \left[\frac{8}{3}, +\infty\right)$ & $\min\{4M,2N\}$ & \cite{Yuan1} \\\hline
4 &  $\left(0,\frac{12}{7}\right] \cup \left[\frac{7}{3}, +\infty\right)$ & $\min\{4M,2N\}$ & \cite{Liu5} \\\hline
4 &  $\left(0,+\infty\right)$ & $\max\{\min\{4M,\frac{12N}{7}\},\min\{\frac{24M}{7},2N\}\}$ & \cite{Wang4} \\\hline
$K>4$ &  $\left(0,\frac{2K^2-2K}{K^2-K+2}\right]$ & $\min\{KM,2N\}$ & \cite{Lee} \\\hline
$K>4$ &  $\left(K-1,+\infty\right]$ & $KM$ & \cite{Mu1} \\\hline
$K>4$ &  $\left(0,\frac{2K^2-2K}{K^2-K+2}\right] \cup \left[\frac{K^2-2K}{K-1}, +\infty\right)$ & $\min\{KM,2N\}$ & \cite{Wang3} \\ \hline
$K>4$ &  $\left(0,\frac{2K^2-2K}{K^2-K+2}\right] \cup \left[\frac{K^2-3K+3}{K-1}, +\infty\right)$ & $\min\{KM,2N\}$ & \cite{Liu5} \\ \hline
$K>4$ &  $\left(\frac{2K^2-2K}{K^2-K+2},\frac{K^2-3K+3}{K-1}\right)$ & unknown & ~ \\ \hline
\end{tabular}
\end{table*}

\textsc{Table} I summarizes the recent advances towards the DoF analysis of the $K$-user MIMO Y channel. In particular, the DoF analysis when $K=3$ and $K=4$ is completed with \cite{Chaaban1} and \cite{Wang4}. The maximum achievable DoF when $K>4$ with the antenna configuration $\frac{N}{M} \in \big(\frac{2K^2-2K}{K^2-K+2}, \frac{K^2-3K+3}{K-1}\big)$ remains unknown.

\subsubsection{Achievable DoF}

\

\textit{Theorem 5}: The achievable DoF for the $K$-user MIMO Y channel at different antenna configurations $\frac{N}{M}$ is given by the union of the single-sided trapezoids characterized by $\left\{Q_1,Q_{\beta} \mid \beta \in \{2,3,4,\cdots,K-2\}\right\}$ shown in \eqref{P_y} at the top of the next page in the DoF plane.

\begin{figure*}[ht]
\hrule
\begin{subequations}\label{P_y}
\begin{align}
&Q_1=\left(\frac{2K^2-2K}{K^2-K+2},\frac{4K^2-4K}{K^2-K+2}\right)\\\label{P_y_beta}
&Q_{\beta}=\left(\beta+\frac{2K(K-1)}{\big(2+K(K-1)-\beta(\beta-1)\big)\binom{K}{\beta}},\frac{4K(K-1)}{2+K(K-1)-\beta(\beta-1)}\right)
\end{align}
\end{subequations}
\hrule
\end{figure*}

\begin{proof}
By \textit{Lemma 2}, to prove this theorem is equivalent to proving the achievability of the points $Q_1$ and $Q_{\beta}$ for $\beta=2,3,\cdots, K-2$. $Q_1$ is proved to be achievable in \cite{Lee}. In what follows, we prove the achievability of each $Q_{\beta}$. Note that the abscissa of $Q_{\beta}$ is located in the interval $(\beta,\beta+1)$.

For the symmetry of each source node, we assume that $d_{i,j}$ for any pair $(i,j)$ is the same and given by $x$, then $d_{total}=K(K-1)x$. For each $\beta$, let the antenna configuration satisfy $\beta M \leq N < (\beta+1) M$.  We choose $q$ out of the $N-\beta M$ basis vectors of the left null space of each ${\cal H}_{\beta}[n]$ as the row vectors of the matrix $\bf P$. In total we can construct $\binom{K}{\beta}q$ row vectors for $\bf P$ as there are $\binom{K}{\beta}q$ different $\beta$-combining channel matrices. On the other hand, the number of the rows of $\bf P$ is $\frac{d_{total}}{2}$. Thus we can let
\begin{align}\label{p_q_dtotal_1}
\frac{d_{total}}{2} = \binom{K}{\beta}q,
\end{align}
which is equivalent to
\begin{align}\label{p_q_dtotal_2}
q =\frac{d_{total}}{2\binom{K}{\beta}}.
\end{align}
Next, we count the number of row vectors in $\bf P$ which are located in the left null space of $\left[{\bf H}_{i,r}~-{\bf H}_{j,r}\right]$. Consider the following $\beta$-combining channel matrix whose span space is
\begin{align}\label{P_i_j_number}
\textrm{span} \left[{\bf H}_{i,r}~{\bf H}_{j,r}~~{\bf H}_{\pi(1),r}~{\bf H}_{\pi(2),r}~\cdots~{\bf H}_{\pi(\beta-2),r}\right]
\end{align}
where $\{\pi(1),\pi(2),\cdots,\pi(\beta-2)\} \subseteq \{1,2,\cdots,K\}\backslash\{i,j\}$.
We can find that there are total $\binom{K-2}{\beta -2}$ different cases. Hence, there are $\binom{K-2}{\beta -2}q$ row vectors which are located in the left null space of $\left[{\bf H}_{i,r}~-{\bf H}_{j,r}\right]$.

From \textit{Theorem 4}, we have
\begin{align}\label{p_constraints_1}
\binom{K-2}{\beta -2}q \geq \frac{d_{total}}{2}-2M+d_{i,j}.
\end{align}
Combining \eqref{p_q_dtotal_2} and \eqref{p_constraints_1}, the relationship between $d_{i,j}$ and $M$ can be derived as
\begin{align}\nonumber
x&=d_{i,j}\\\nonumber
& \leq 2M-\frac{d_{total}}{2}+\frac{d_{total}\binom{K-2}{\beta -2}}{2\binom{K}{\beta}}\\\label{x_y}
&= 2M-\frac{K(K-1)x}{2}+\frac{K(K-1)x\binom{K-2}{\beta -2}}{2\binom{K}{\beta}}
\end{align}
From \eqref{x_y}, we obtain that
\begin{align}\nonumber
x &\leq \frac{4\binom{K}{\beta}M}{2\binom{K}{\beta}+K(K-1)\binom{K}{\beta}-K(K-1)\binom{K-2}{\beta-2}}\\\nonumber
&= \frac{4\binom{K}{\beta}M}{2\binom{K}{\beta}+K(K-1)\binom{K}{\beta}-K(K-1)\big[\binom{K}{\beta}\frac{\beta (\beta-1)}{K(K-1)}\big]}\\
&=\frac{4M}{2+K(K-1)-\beta (\beta-1)}.
\end{align}
Thus, the maximum achievable DoF is
\begin{equation}
d_{total}=K(K-1)x=\frac{4K(K-1)M}{2+K(K-1)-\beta (\beta-1)}.
\end{equation}
On the other hand, since $q \leq N-\beta M$, by considering \eqref{p_q_dtotal_2} we have $N \geq \beta M+\frac{2K(K-1)M}{\big(2+K(K-1)-\beta(\beta-1)\big)\binom{K}{\beta}}$.
Hence, the achievabiliy of the point $Q_{\beta}$ in the DoF plane is proved.
\end{proof}

From \textit{Theorem 5}, we can express the achievable DoF explicitly at certain antenna configuration regions $\frac{N}{M}$ as
\begin{align}\label{dof_KY_a}
d_{total} = \left\{
                    \begin{array}{ll}
                      2N, & \hbox{$\frac{N}{M} \in \left(0, \frac{2K^2-2K}{K^2-K+2}\right]$,} \\
                      \frac{(4K^2-4K)M}{K^2-K+2}, & \hbox{$\frac{N}{M} \in \left(\frac{2K^2-2K}{K^2-K+2}, 2\right]$,} \\
                      \frac{(2K^2-2K)N}{K^2-K+2}, & \hbox{$\frac{N}{M} \in \left(2, 2+\frac{4}{K(K-1)}\right]$,} \\
                      \vdots\\
                      \frac{K(K-1)N}{K^2-3K+3}, & \hbox{$\frac{N}{M} \in \left(\gamma, \frac{K^2-3K+3}{K-1}\right]$,} \\
                      KM, & \hbox{$\frac{N}{M} \in \left(\frac{K^2-3K+3}{K-1}, +\infty \right)$,}
                    \end{array}
                  \right.
\end{align}
where $\gamma$ is some value between $\left(2+\frac{4}{K(K-1)},K-2\right)$ which does not have explict expression but can be computed numerically.

\begin{figure}[t]
\begin{centering}
\includegraphics[scale=0.48]{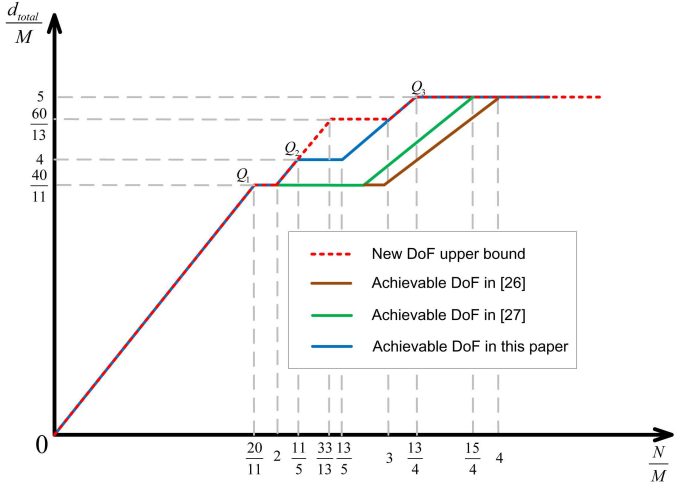}
\vspace{-0.1cm}
 \caption{New DoF upper bound and its achievability for 5-user MIMO Y channel.}\label{DoF_Y_5}
\end{centering}
\vspace{-0.2cm}
\end{figure}

Comparing with the upper bound in \textit{Theorem 1}, we find that the DoF upper bound under the antenna configuration $\frac{N}{M} \in \big(0, 2+\frac{4}{K(K-1)}\big] \cup \big[K-2, +\infty\big)$ is tight by GSA when $K>4$. By comparing with the previous results summarized in Table I, it is seen that the unknown region for $\frac{N}{M}$ to achieve the maximum DoF is reduced to $\left(2+\frac{4}{K(K-1)},K-2\right)$. Fig. \ref{DoF_Y_5} illustrates the new DoF upper bound and its achievability when $K=5$.

In the case with $K=4$, the achievable DoF is the same as the DoF upper bound for all $\frac{N}{M}$. This is consistent with the results obtained in \cite{Wang4}.

\begin{figure}[t]
\begin{centering}
\includegraphics[scale=0.6]{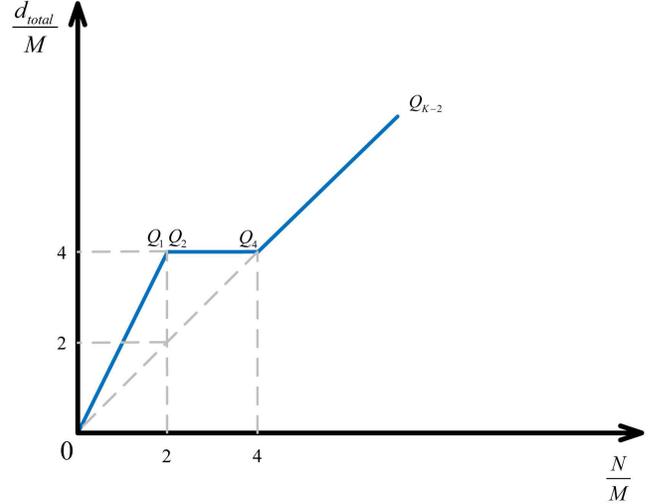}
\vspace{-0.1cm}
 \caption{Asymptotic DoF behavior when $K \rightarrow \infty$ for $K$-user MIMO Y channel.}\label{DoF_Y_K_infinity}
\end{centering}
\vspace{-0.2cm}
\end{figure}

\textit{Corollary 1}: When $K \rightarrow \infty$, the achievable DoF for the $K$-user MIMO Y channel is
\begin{align}\label{dof_Y_K_infinity}
d_{total} = \left\{
                    \begin{array}{ll}
                      2N, & \hbox{$\frac{N}{M} \in \big(0, 2\big]$,} \\
                      4M, & \hbox{$\frac{N}{M} \in \big(2,4\big]$,} \\
                      N, & \hbox{$\frac{N}{M} \in \big(4, +\infty \big)$,}
                    \end{array}
                  \right.
\end{align}
as illustrated in Fig. \ref{DoF_Y_K_infinity}.
\begin{proof}
According to \textit{Theorem 5}, we have $Q_1=Q_2=(2,4)$, $Q_3=(3,4)$ and $Q_4=(4,4)$ when $K \rightarrow \infty$. Then we have
\begin{equation}
\lim \limits_{K \rightarrow \infty} Q_{K-2}=\lim \limits_{K \rightarrow \infty} \left(\frac{K^2-3K+3}{K-1},K\right)=\lim \limits_{K \rightarrow \infty} \left(K,K\right).
\end{equation}
The slope of the line from $0$ to $Q_1$ $(0 \rightarrow Q_1)$ is two while the slope of $0\rightarrow Q_{K-2}$ is one. Next, we show that the slope of $0\rightarrow Q_{\beta}$ is less than one for any $\beta \in \{5,6,\cdots,K-3\}$. Denote $l_{\beta}$ as the slope of $0\rightarrow Q_{\beta}$ and we have
\begin{align}\nonumber
l_{\beta}&=\frac{4K(K-1)}{\beta\big(2+K(K-1)-\beta(\beta-1)\big)\binom{K}{\beta}+2K(K-1)}\\\nonumber
&\leq \frac{4K(K-1)}{\beta\big(2+K(K-1)-\beta(\beta-1)\big)\binom{K}{3}+2K(K-1)}\\\nonumber
&= \frac{4K(K-1)}{\left[\beta\big(2+K(K-1)-\beta(\beta-1)\big)\frac{K-2}{6}+2\right]K(K-1)}\\\nonumber
&< \frac{4K(K-1)}{\left[2+2\right](K-1)}\\
&=1.
\end{align}
This indicates that $\left\{Q_{\beta} \mid \beta \in \{5,6,\cdots,K-3\}\right\}$ are all located in the single-sided trapezoids characterized by the point $Q_{K-2}$.
The corollary is thus proved.
\end{proof}

\subsubsection{An example when DoF of $KM$ is achievable}

\

Here, we illustrate the specific construction of the compression matrix $\bf P$ and the source precoding matrix ${\bf V}_{i,j}$ to achieve the DoF $KM$ when $\frac{N}{M} \geq \frac{K^2-3K+3}{K-1}$. We let $d_{i,j}$ be $\frac{M}{K-1}$ for all $i \neq j$. The data switch matrix $\bf D$ is
\begin{equation}\label{D_Y_2}
{{\bf D}}=\left[\begin{array}{ccccc}
                 0 & \frac{M}{K-1} & \cdots & \frac{M}{K-1} & \frac{M}{K-1} \\
                 \frac{M}{K-1} & 0 & \cdots & \frac{M}{K-1} & \frac{M}{K-1} \\
                 \vdots & \vdots & \ddots & \vdots & \vdots \\
                 \frac{M}{K-1} & \frac{M}{K-1} & \cdots & 0 & \frac{M}{K-1} \\
                 \frac{M}{K-1} & \frac{M}{K-1} & \cdots & \frac{M}{K-1} & 0
               \end{array}
\right].
\end{equation}
We separate the analysis into two cases.

\textbf{Case 1: $M$ is divisible by $K-1$.} Denote
\begin{equation}\label{desired_Y}
\textbf{s}_{\oplus}=\left[\begin{array}{c}
                            {\bf s}_{1,2}+{\bf s}_{2,1} \\
                            {\bf s}_{1,3}+{\bf s}_{3,1} \\
                            \vdots \\
                            {\bf s}_{i,j}+{\bf s}_{j,i} \\
                            \vdots \\
                            {\bf s}_{K-1,K}+{\bf s}_{K,K-1}
                          \end{array}
\right]
\end{equation}
as the network-coded symbol vector expected to obtain at the relay, where each ${\bf s}_{i,j}$ is a $\frac{M}{K-1} \times 1$ vector.

Let the $\frac{KM}{2} \times N$ compression matrix $\bf P$ be stacked by $\frac{M}{K-1}\times N$ submatrices ${\bf P}_{i,j}$ by row. We design the compression matrix $\bf P$ by the method in the proof of \textit{Theorem 5}. Each submatrix ${\bf P}_{i,j}$ is designed as \eqref{Y_null} at the top of the next page.
\begin{figure*}[ht]
\begin{equation}\label{Y_null}
{\bf P}_{i,j}^T \subseteq \textbf{Null}~\big[{\bf H}_{1,r}~\cdots~{\bf H}_{i-1,r}~{\bf H}_{i+1,r}~\cdots~{\bf H}_{j-1,r}~{\bf H}_{j+1,r}~\cdots~{\bf H}_{K,r}\big]^T.
\end{equation}
\hrule
\end{figure*}
Then we design the precoding matrices ${\bf V}_{i,j}$ for each source node. Each pair of $M \times \frac{M}{K-1}$ precoding matrices is designed as
\begin{align}\label{GSA_v_Y}
\left[\begin{array}{c}
        {\bf V}_{i,j} \\
        {\bf V}_{j,i}
      \end{array}
\right] \subseteq \textbf{Null}~\left[{\bf P}{\bf H}_{i,r}~-{\bf P}{\bf H}_{j,r}\right].
\end{align}

Similar to \eqref{b_12_Y_4}, we can obtain the direction of the aligned signals of signal pair (1, 2) as
\begin{align}
{\bf B}_{1,2}={\bf P}{\bf H}_{1,r}{\bf V}_{1,2}={\bf P}{\bf H}_{3,r}{\bf V}_{3,1}=\left[\begin{array}{ccccccc}
\alpha_{1,2}^1 & \cdots & 0 \\
0 & \cdots & 0 \\
\vdots & \ddots & \vdots \\
0 & \cdots & \alpha_{1,2}^{d_{1,2}}\\
0 & \cdots & 0\\
\vdots & \ddots & \vdots\\
0 & \cdots & 0\\
\end{array}\right],
\end{align}
where $\alpha_{1,2}^l$ ($1 \leq l \leq d_{1,2}$) is a constant and ${\bf B}_{1,2}$ is a $\frac{KM}{2} \times d_{1,2}$ matrix.

Plugging \eqref{Y_null} and \eqref{GSA_v_Y} into \eqref{y_r_projection}, we can obtain the signals after compression as
\begin{align}\label{y_r_projection_Y}
\hat{{\bf y}}_r={\bm \alpha} \textbf{s}_{\oplus} + {\bf P}{\bf n}_r
\end{align}
where ${\bm \alpha}$ is a diagonal matrix. Then, the network-coded symbol vector can be readily estimated from \eqref{y_r_projection_Y}.

During the BC phase, we use the method of interference nulling to design the precoding matrix {\bf U}. We can write {\bf U} as follows.
\begin{equation}\label{UU}
{\bf U}=
\left[
\begin{array}{ccc}
 {\bf U}_{1}~{\bf U}_{2}~\cdots~{\bf U}_{\frac{K(K-1)}{2}}
\end{array}
\right],
\end{equation}
where each ${\bf U}_i$ is an $N \times \frac{M}{K-1}$ matrix and
\begin{eqnarray}\nonumber\label{UU3}
{\bf U}_1&\subseteq \textbf{Null}~\big[{\bf G}_{r,3}^T~{\bf G}_{r,4}^T~\cdots~{\bf G}_{r,K}^T\big]^T~\\\nonumber
{\bf U}_2&\subseteq \textbf{Null}~\big[{\bf G}_{r,2}^T~{\bf G}_{r,4}^T~\cdots~{\bf G}_{r,K}^T\big]^T~\\\nonumber
\vdots\\\nonumber
\vdots\\
{\bf U}_{\frac{K(K-1)}{2}}&\subseteq \textbf{Null}~\big[{\bf G}_{r,1}^T~{\bf G}_{r,2}^T~\cdots~{\bf G}_{r,K-2}^T\big]^T
\end{eqnarray}
The matrix ${\bf U}_i$ ($N \times \frac{M}{K-1}$) exists if and only if $N-(K-2)M \geq \frac{M}{K-1}$, or equivalently $N \geq \frac{(K^2-3K+3)M}{K-1}$. Hence, we can apply GSA-based transmission scheme when $M$ is divisible by $K-1$ and $N \geq \frac{(K^2-3K+3)M}{K-1}$ to achieve the DoF upper bound $KM$.

\textbf{Case 2: $M$ is not divisible by $K-1$.} In this case, we use the idea of the symbol extension \cite{Jafar1} together with GSA to prove the achievability of the DoF upper bound $KM$. We consider the $(K-1)$-symbol extension of the channel model, where the channel coefficients do not necessarily vary over time. The received signal at the relay can be written as
\begin{align}\nonumber
{\bf y}_r&=\left[\begin{array}{c}{\bf y}_r(1)\\{\bf y}_r(2)\\\vdots\\{\bf y}_r(K-1)\end{array}\right]\\\nonumber
&=\left[\begin{array}{cccc}
{\bf H}(1) & {\bf 0} & \cdots & {\bf 0} \\
{\bf 0} & {\bf H}(2) & \cdots & {\bf 0} \\
\vdots & \vdots & \ddots & {\bf 0} \\
{\bf 0} & {\bf 0} & \cdots & {\bf H}(K-1)
\end{array}
\right]\left[\begin{array}{c}{\bf x}(1)\\{\bf x}(2)\\\vdots\\{\bf x}(K-1)\end{array}\right]\\\nonumber
&~~~~+\left[\begin{array}{c}{\bf n}_r(1)\\{\bf n}_r(2)\\\vdots\\{\bf n}_r(K-1)\end{array}\right]\\\label{y_r_symbol_Y}
&={\bf H}^{\S}{\bf x}^{\S}+{\bf n}_r^{\S}.
\end{align}
where ${\bf y}_r(t)$, ${\bf H}(t)$, ${\bf x}(t)$ and ${\bf n}_r(t)$ denote the $t$-th time slot of received signals, channel matrices, transmitted signals and noise, ${\bf H}^{\S}$ denotes the equivalent channel matrix, ${\bf x}^{\S}$ denotes the equivalent transmitted signals, and ${\bf n}_r^{\S}$ denotes the equivalent noise.

Note that ${\bf H}^{\S}$ is a $(K-1)N\times(K-1)KM$ matrix. The system model is equivalent to the $K$-user MIMO Y channel with each source node equipped with $(K-1)M$ antennas and the relay equipped with $(K-1)N$ antennas. It turns to be \textbf{Case 1} and we can then apply GSA to achieve the DoF $(K-1)KM$ over $(K-1)$ channel uses. This implies that the DoF of $KM$ per channel use is achievable in the original $K$-user MIMO Y channel. The antenna constraint can be written as
\begin{equation}\nonumber
(K-1)N-(K-2)(K-1)M \geq \frac{(K-1)M}{K-1}
\end{equation}
or equivalently,
\begin{equation}\label{s_NKMYYYY}
N \geq \frac{(K^2-3K+3)M}{K-1}.
\end{equation}

The above analysis shows that the generalized signal alignment based transmission scheme can achieve the DoF of $KM$ when $N \geq \frac{(K^2-3K+3)M}{K-1}$ in the $K$-user MIMO Y channel.

\subsection{Multi-pair MIMO two-way relay channel}
The multi-pair MIMO two-way relay channel consists of $\frac{K}{2}$ pairs of source nodes, each source node equipped with $M$ antennas, and one relay node, equipped with $N$ antennas. The two source nodes, denoted as $i$ and $K+1-i$, exchange messages with each other with the help of the relay. Previous studies \cite{Tian,Tian1} analyzed the achievable DoF for this model. The maximum achievable DoF with the antenna configuration $\frac{N}{M} \in \big(\frac{2K}{K+2}, K\big)$ remains unknown. Our result is given in the following theorem.

\subsubsection{Achievable DoF}

\

\textit{Theorem 6}:  The achievable DoF for the multi-pair MIMO two-way relay channel at different antenna configurations $\frac{N}{M}$ is given by the union of the single-sided trapezoids characterized by the following points in the DoF plane:
\begin{subequations}\label{P_MP}
\begin{align}
&Q_1=\left(\frac{2K}{K+2},\frac{4K}{K+2}\right)\\\label{P_MP_beta}
&Q_{\beta}=\left(\beta+\frac{2K}{\big(2+K-\beta\big)\binom{K/2}{\beta/2}},\frac{4K}{2+K-\beta}\right)
\end{align}
\end{subequations}
where $\beta \in \{2,4,\cdots,K-2\}$. Note that $\beta$ is even.
\begin{proof}
By \textit{Lemma 2}, to prove this theorem is equivalent to proving the achievability of the points $Q_1$ and $Q_{\beta}$ for $\beta=2,3,\cdots, K-2$. $Q_1$ is proved to be achievable in \cite{Tian} with SA. In what follows, we prove the achievability of each $Q_{\beta}$. Note that the abscissa of $Q_{\beta}$ is located in the interval $(\beta,\beta+1)$.

For the symmetry of each source node, we assume that $d_{i,j}$ for any pair $(i,j)$ is the same and given by $x$, then $d_{total}=Kx$. For each $\beta$, let the antenna configuration satisfy $\beta M \leq N < (\beta+1) M$. Different from the method of the design of $\bf{P}$ in the previous subsection, we define the paired-combining channel matrix $[{\bf H}_{i,r}~{\bf H}_{K+1-i,r}]$, where $i \in \{1,2,\cdots, \frac{K}{2}\}$. We only choose $q$ out of the $N-\beta M$ basis vectors of the left null space of each ${\cal H}_{\beta}[n]$, whose span space consists of $\frac{\beta}{2}$ paired-combining channel matrices, as the row vectors of the matrix $\bf P$. In total we can construct $\binom{K/2}{\beta/2}q$ row vectors for $\bf P$ as there are $\binom{K/2}{\beta/2}q$ different $\beta$-combining channel matrices. On the other hand, the number of the rows of $\bf P$ is $\frac{d_{total}}{2}$. Thus we can let
\begin{align}\label{p_q_dtotal_1_MP}
\frac{d_{total}}{2} = \binom{K/2}{\beta/2}q,
\end{align}
which is equivalent to
\begin{align}\label{p_q_dtotal_2_MP}
q = \frac{d_{total}}{2\binom{K/2}{\beta/2}}.
\end{align}
Next, we count the number of row vectors in $\bf P$ which are located in the left null space of $\left[{\bf H}_{i,r}~-{\bf H}_{j,r}\right]$. Consider the following $\beta$-combining channel matrix whose span space is
\begin{align}\label{P_i_j_number_MP}
\textrm{span} \left[{\bf H}_{i,r}~{\bf H}_{j,r}~~{\bf H}_{\pi(1),r}~{\bf H}_{\pi(2),r}~\cdots~{\bf H}_{\pi(\beta-2),r}\right]
\end{align}
where $\left\{\pi(1),\pi(2),\cdots,\pi\left(\frac{\beta-2}{2}\right)\right\} \subseteq \{1,2,\cdots,\frac{K}{2}\}\backslash\{i,j\}$ and $\pi\left(k+\frac{\beta-2}{2}\right)=K+1-\pi\left(k\right)$, $k=1,2,\cdots,\frac{\beta-2}{2}$.
We can find that there are total $\binom{K/2-1}{\beta/2-1}$ different cases. Hence, there are $\binom{K/2-1}{\beta/2-1}q$ row vectors which are located in the left null space of $\left[{\bf H}_{i,r}~-{\bf H}_{j,r}\right]$.

From \textit{Theorem 4}, we have
\begin{align}\label{p_constraints_1_MP}
\binom{K/2-1}{\beta/2-1}q \geq \frac{d_{total}}{2}-2M+d_{i,j}.
\end{align}
Combining \eqref{p_q_dtotal_2_MP} and \eqref{p_constraints_1_MP}, the relationship between $d_{i,j}$ and $M$ can be derived as
\begin{align}\nonumber
x&=d_{i,j} \\\nonumber
&\leq 2M-\frac{d_{total}}{2}+\frac{d_{total}\binom{K/2-1}{\beta/2-1}}{2\binom{K/2}{\beta/2}}\\\label{x_MP}
&= 2M-\frac{Kx}{2}+\frac{\beta x}{2}
\end{align}
From \eqref{x_MP}, we obtain that
\begin{align}
x & \leq \frac{4M}{2+K-\beta}.
\end{align}
Thus, the maximum achievable DoF is
\begin{equation}
d_{total}=Kx=\frac{4KM}{2+K-\beta}
\end{equation}
On the other hand, since $q \leq N-\beta M$, by considering \eqref{p_q_dtotal_2_MP} we have $N \geq \beta M+\frac{2KM}{\big(2+K-\beta\big)\binom{K/2}{\beta/2}}$.
Hence, the achievabiliy of the point $Q_{\beta}$ in the DoF plane is proved.
\end{proof}

From \textit{Theorem 6}, we can express the achievable DoF explicitly at certain antenna configuration regions $\frac{N}{M}$ as
\begin{align}\label{dof_KY_a}
d_{total} = \left\{
                    \begin{array}{ll}
                      2N, & \hbox{$\frac{N}{M} \in \left(0, \frac{2K}{K+2}\right]$,} \\
                      \frac{4KM}{K+2}, & \hbox{$\frac{N}{M} \in \left(\frac{2K}{K+2}, 2\right]$,} \\
                      \frac{2KN}{K+2}, & \hbox{$\frac{N}{M} \in \left(2, 2+\frac{4}{K}\right]$,} \\
                      \vdots\\
                      \frac{KN}{K-1}, & \hbox{$\frac{N}{M} \in \left(\gamma, K-1\right]$,} \\
                      KM, & \hbox{$\frac{N}{M} \in \left(K-1, +\infty \right)$,}
                    \end{array}
                  \right.
\end{align}
where $\gamma$ is some value between $\left(2+\frac{4}{K},K-2\right)$ which does not have explict expression but can be computed numerically.

Comparing with the upper bound in \textit{Theorem 2}, we find that the DoF upper bound under the antenna configuration $\frac{N}{M} \in \big(0, 2+\frac{4}{K}\big] \cup \big[K-2, +\infty\big)$ is tight by GSA when $K>4$. By comparing with the previous results in \cite{Tian1,Tian}, it is seen that the unknown region for $\frac{N}{M}$ to achieve the maximum DoF is reduced to $\left(2+\frac{4}{K},K-2\right)$.

If $K=4$, then only consider $\beta =2$ and the achievable DoF is the same as the DoF upper bound for all $\frac{N}{M}$.

\textit{Corollary 2}: When $K \rightarrow \infty$, the achievable DoF for the multi-pair MIMO two-way relay channel is
\begin{equation}\label{dof_MP_K_infinity}
d_{total} = \left\{
                    \begin{array}{ll}
                      2N, & \hbox{$\frac{N}{M} \in \big(0, 2\big]$,} \\
                      4M, & \hbox{$\frac{N}{M} \in \big(2,4\big]$,} \\
                      N, & \hbox{$\frac{N}{M} \in \big(4, +\infty \big)$.}
                    \end{array}
                  \right.
\end{equation}
\begin{proof}
The proof is similar to the proof of \textit{Corollary 1} and hence omitted.
\end{proof}

\subsubsection{An example when DoF of $KM$ is achievable}

\

Here, we illustrate the specific construction of the compression matrix $\bf P$ and the source precoding matrix ${\bf V}_{i,j}$ to achieve the DoF $KM$ when $\frac{N}{M} \geq K-1$. We let $d_{i,j}$ be $M$ for all $i \neq j$. The data switch matrix $\bf D$ is

\begin{equation}\label{D_MP}
{{\bf D}}=\left[\begin{array}{ccccc}
                 0 & 0 & \cdots & 0 & M \\
                 0 & 0 & \cdots & M & 0 \\
                 \vdots & \vdots & \ddots & \vdots & \vdots \\
                 0 & M & \cdots & 0 & 0 \\
                 M & 0 & \cdots & 0 & 0
               \end{array}
\right].
\end{equation}

Let the $\frac{KM}{2} \times N$ compression matrix $\bf P$ be stacked by $M \times N$ submatrices ${\bf P}_{i,j}$ by row. We design the compression matrix $\bf P$ by the method in the proof of \textit{Theorem 6}. Each submatrix ${\bf P}_{i,j}$ is designed as \eqref{MP_null}.
\begin{figure*}[ht]
\begin{equation}\label{MP_null}
{\bf P}_{i,\bar{i}}^T \subseteq \textbf{Null}~\big[{\bf H}_{1,r}~\cdots~{\bf H}_{i-1,r}~{\bf H}_{i+1,r}~\cdots~{\bf H}_{\bar{i}-1,r}~{\bf H}_{\bar{i}+1,r}~\cdots~{\bf H}_{K,r}\big]^T.
\end{equation}
\hrule
\end{figure*}
Then we design the precoding matrices ${\bf V}_{i,j}$ for each source node. Each source pair of source precoding matrices is designed as
\begin{align}\label{GSA_v_MP}
\left[\begin{array}{c}
        {\bf V}_{i,\bar{i}} \\
        {\bf V}_{\bar{i},i}
      \end{array}
\right] \subseteq \textbf{Null}~\left[{\bf P}{\bf H}_{i,r}~-{\bf P}{\bf H}_{\bar{i},r}\right].
\end{align}
Plugging \eqref{MP_null} and \eqref{GSA_v_MP} into \eqref{y_r_projection}, we can obtain the signals after compression as
\begin{align}\label{y_r_projection_MP}
\hat{{\bf y}}_r={\bm \alpha} \textbf{s}_{\oplus} + {\bf P}{\bf n}_r
\end{align}
where ${\bm \alpha}$ is a diagonal matrix and $\textbf{s}_{\oplus}=[{\bf s}_{1,K}^T+{\bf s}_{K,1}^T, {\bf s}_{2,K-1}^T+{\bf s}_{K-1,2}^T, \cdots, {\bf s}_{i,\bar{i}}^T+{\bf s}_{\bar{i},i}^T, \cdots, {\bf s}_{\frac{K}{2},\frac{K}{2}+1}^T+{\bf s}_{\frac{K}{2}+1,\frac{K}{2}}^T]^T$. Then, the network-coded symbol vector can be readily estimated from \eqref{y_r_projection_MP}.

During the BC phase, we use similar interference nulling method as in the $K$-user MIMO Y channel to design the precoding matrix {\bf U}. We can write {\bf U} as follows.
\begin{equation}\label{UU_MP}
{\bf U}=
\left[
\begin{array}{ccc}
 {\bf U}_{1}~{\bf U}_{2}~\cdots~{\bf U}_{\frac{K}{2}}
\end{array}
\right]
\end{equation}
where each ${\bf U}_i$ is an $N \times M$ matrix and
\begin{align}\nonumber\label{UU3_MP}
{\bf U}_1&\subseteq \textbf{Null}~\big[{\bf G}_{r,2}^T~{\bf G}_{r,3}^T~\cdots~{\bf G}_{r,K-1}^T\big]^T~\\\nonumber
{\bf U}_2&\subseteq \textbf{Null}~\big[{\bf G}_{r,1}^T~{\bf G}_{r,3}^T~\cdots~{\bf G}_{r,K-2}^T~{\bf G}_{r,K}^T\big]^T~\\\nonumber
\vdots\\\nonumber
\vdots\\
{\bf U}_{\frac{K}{2}}&\subseteq \textbf{Null}~\big[{\bf G}_{r,1}^T~{\bf G}_{r,2}^T~\cdots~{\bf G}_{r,\frac{K}{2}-1}^T~{\bf G}_{r,\frac{K}{2}+2}^T~\cdots~{\bf G}_{K}^T\big]^T
\end{align}
We can see that ${\bf U}_i$ ($N \times M$) exists if and only if
\begin{equation}\label{s_NKMYYY_MP}
N-(K-2)M \geq M
\end{equation}
or equivalently,
\begin{equation}\label{s_NKMYYYY_MP}
N \geq (K-1)M.
\end{equation}

The following steps are similar to those in the previous subsection.

\subsection{Generalized MIMO two-way X relay channel}
The generalized MIMO two-way X relay channel consists of two groups of source nodes of size $\frac{K}{2}$, each equipped with $M$ antennas, and one relay node, equipped with $N$ antennas. Each source node in one group, denoted as $i=1,~2,~\cdots,~\frac{K}{2}$, exchanges independent messages with every source node in the other group, denoted as $i=\frac{K}{2}+1,~\frac{K}{2}+2,~\cdots,~K$, with the help of the relay. In the special case of $K=4$, i.e. MIMO two-way X relay channel, the work in \cite{Xiang} showed that the DoF of $2N$ is achievable when $N \leq \lfloor\frac{8M}{5}\rfloor$; the work in \cite{Liu3} showed that the DoF of $4M$ is achievable when $N \geq \lceil\frac{5M}{2}\rceil$. Our result is given in the following theorem.

\subsubsection{Achievable DoF}

\

\textit{Theorem 7}:  The achievable DoF for the generalized MIMO two-way X relay channel at different antenna configurations $\frac{N}{M}$ is given by the union of the single-sided trapezoids characterized by the following points in the DoF plane:
\begin{subequations}\label{P_X}
\begin{align}
&Q_1=\left(\frac{2K^2}{K^2+4},\frac{4K^2}{K^2+4}\right)\\\label{P_X_beta}
&Q_{\beta}=\left(\beta+\frac{2K^2}{\big(4+K^2-\beta^2\big)\binom{K/2}{\beta/2}\binom{K/2}{\beta/2}},\frac{4K^2}{4+K^2-\beta^2}\right)
\end{align}
\end{subequations}
where $\beta \in \{2,4,\cdots,K-2\}$. Note that $\beta$ is even.
\begin{proof}
By \textit{Lemma 2}, to prove this theorem is equivalent to proving the achievability of the points $Q_1$ and $Q_{\beta}$ for $\beta=2,3,\cdots, K-2$. $Q_1$ is achievable with SA, we omit the detailed proof here. In what follows, we mainly prove the DoF achievability of $Q_{\beta}$ when $\beta \geq 2$. Note that the abscissa of $Q_{\beta}$ is located in the interval $(\beta,\beta+1)$.

For the symmetry of each source node, we assume that $d_{i,j}$ for any pair $(i,j)$ is the same and given by $x$, then $d_{total}$ is $\frac{K^2}{2}x$. For each $\beta$, let the antenna configuration satisfy $\beta M \leq N < (\beta+1) M$. Define ${\cal H}_{group~1}=\{{\bf H}_{1,r},{\bf H}_{2,r},\cdots,{\bf H}_{\frac{K}{2},r}\}$ and ${\cal H}_{group~2}=\{{\bf H}_{\frac{K}{2}+1,r},{\bf H}_{\frac{K}{2}+2,r},\cdots,{\bf H}_{K,r}\}$. We only choose $q$ out of the $N-\beta M$ basis vectors of the left null space of each ${\cal H}_{\beta}[n]$, whose span space consists of $\frac{\beta}{2}$ channel matrices in ${\cal H}_{group~1}$ and $\frac{\beta}{2}$ channel matrices in ${\cal H}_{group~2}$, as the row vectors of the matrix $\bf P$. In total we can construct $\binom{K/2-1}{\beta/2-1}\binom{K/2-1}{\beta/2-1}q$ row vectors for $\bf P$ as there are $\binom{K/2-1}{\beta/2-1}\binom{K/2-1}{\beta/2-1}$ different $\beta$-combining channel matrices. On the other hand, the number of the rows of $\bf P$ is $\frac{d_{total}}{2}$. Thus we can let
\begin{align}\label{p_q_dtotal_1_X}
\frac{d_{total}}{2} =\binom{K/2}{\beta/2}\binom{K/2}{\beta/2}q,
\end{align}
which is equivalent to
\begin{align}\label{p_q_dtotal_2_X}
q = \frac{d_{total}}{2\binom{K/2}{\beta/2}\binom{K/2}{\beta/2}}.
\end{align}
Next, we count the number of row vectors in $\bf P$ which are located in the left null space of $\left[{\bf H}_{i,r}~-{\bf H}_{j,r}\right]$. Consider the following $\beta$-combining channel matrix whose span space can be expressed as \eqref{P_i_j_number_X} at the top of the next page.
\begin{figure*}[ht]
\begin{align}\label{P_i_j_number_X}
\textrm{span} \left[{\bf H}_{i,r}~~{\bf H}_{j,r}~~\underbrace{{\bf H}_{\pi(1),r}~~{\bf H}_{\pi(2),r}~~\cdots~~{\bf H}_{\pi\left(\frac{\beta-2}{2}\right),r}}_{\textrm{in}~{\cal H}_{group~1}} ~~\underbrace{{\bf H}_{\pi\left(\frac{\beta-2}{2}+1\right),r}~~{\bf H}_{\pi\left(\frac{\beta-2}{2}+2\right),r}~~\cdots~~{\bf H}_{\beta-2,r}}_{\textrm{in}~{\cal H}_{group~2}}\right]
\end{align}
\hrule
\end{figure*}
We can find that there are total $\binom{K/2-1}{\beta/2-1}\binom{K/2-1}{\beta/2-1}$ different cases. Hence, there are $\binom{K/2-1}{\beta/2-1}\binom{K/2-1}{\beta/2-1}q$ row vectors which are located in the left null space of $\left[{\bf H}_{i,r}~-{\bf H}_{j,r}\right]$.

From \textit{Theorem 4}, we have
\begin{align}\label{p_constraints_1_X}
\binom{K/2-1}{\beta/2-1}\binom{K/2-1}{\beta/2-1}q \geq \frac{d_{total}}{2}-2M+d_{i,j}.
\end{align}
Combining \eqref{p_q_dtotal_2_X} and \eqref{p_constraints_1_X}, the relationship between $d_{i,j}$ and $M$ can be derived as
\begin{align}\nonumber
x&=d_{i,j}\\\nonumber
& \leq 2M-\frac{d_{total}}{2}+\frac{d_{total}\binom{K/2-1}{\beta/2-1}\binom{K/2-1}{\beta/2-1}} {2\binom{K/2}{\beta/2}\binom{K/2}{\beta/2}}\\\label{x_X}
&= 2M-\frac{K^2x}{4}+\frac{\beta^2 x}{4}
\end{align}
From \eqref{x_X}, we obtain that
\begin{align}
x & \leq \frac{8M}{4+K^2-\beta^2}.
\end{align}
Thus, the maximum achievable DoF is
\begin{equation}
d_{total}=\frac{K^2}{2}x=\frac{4K^2 M}{4+K^2-\beta^2}
\end{equation}
On the other hand, since $q \leq N-\beta M$, by considering \eqref{p_q_dtotal_2_X} we have $N \geq \beta M+\frac{2K^2 M}{\big(4+K^2-\beta^2\big)\binom{K/2}{\beta/2}\binom{K/2}{\beta/2}}$.
Hence, the achievabiliy of the point $Q_{\beta}$ in the DoF plane is proved.
\end{proof}

From \textit{Theorem 7}, we can express the achievable DoF explicitly at certain antenna configuration regions $\frac{N}{M}$ as
\begin{align}\label{dof_KX_a}
d_{total} = \left\{
                    \begin{array}{ll}
                      2N, & \hbox{$\frac{N}{M} \in \left(0, \frac{2K^2}{K^2+4}\right]$,} \\
                      \frac{4K^2 M}{K^2+4}, & \hbox{$\frac{N}{M} \in \left(\frac{2K^2}{K^2+4}, 2\right]$,} \\
                      \frac{2K^2 N}{K^2+4}, & \hbox{$\frac{N}{M} \in \left(2, 2+\frac{8}{K^2}\right]$,} \\
                      \vdots\\
                      \frac{K^2 N}{K^2-2K+2}, & \hbox{$\frac{N}{M} \in \left(\gamma, \frac{K^2-2K+2}{K}\right]$,} \\
                      KM, & \hbox{$\frac{N}{M} \in \left(\frac{K^2-2K+2}{K}, +\infty \right)$,}
                    \end{array}
                  \right.
\end{align}
where $\gamma$ is some value between $\left(2+\frac{8}{K^2},K-2\right)$ which does not have explict expression but can be computed numerically.

Comparing with the upper bound in \textit{Theorem 3}, we find that the DoF upper bound under the antenna configuration $\frac{N}{M} \in \big(0, 2+\frac{8}{K^2}\big] \cup \big[K-2, +\infty\big)$ is tight by GSA when $K>4$.

In the case with $K=4$, the achievable DoF is the same as the DoF upper bound for all $\frac{N}{M}$. This is consistent with the results obtained in \cite{Wang4}.

\textit{Corollary 3}: When $K \rightarrow \infty$, the achievable DoF for the generalized MIMO two-way X relay channel is
\begin{align}\label{dof_X_K_infinity}
d_{total} = \left\{
                    \begin{array}{ll}
                      2N, & \hbox{$\frac{N}{M} \in \big(0, 2\big]$,} \\
                      4M, & \hbox{$\frac{N}{M} \in \big(2,4\big]$,} \\
                      N, & \hbox{$\frac{N}{M} \in \big(4, +\infty \big)$.}
                    \end{array}
                  \right.
\end{align}
\begin{proof}
The proof is similar to the proof of \textit{Corollary 1} and hence omitted.
\end{proof}

\subsubsection{An example when DoF of $KM$ is achievable}

\

Here, we illustrate the specific construction of the compression matrix $\bf P$ and the source precoding matrix ${\bf V}_{i,j}$ to achieve the DOF $KM$ when $\frac{N}{M} \geq \frac{K^2-2K+2}{K}$. We let $d_{i,j}$ be $\frac{2M}{K}$ for all $i \neq j$. The data switch matrix $\bf D$ is
\begin{equation}\label{D_X_1}
{{\bf D}}=\left[\begin{array}{cccccc}
                  0 & \cdots & 0 & d_{1,\frac{K}{2}+1} & \cdots & d_{1,K} \\
                  \vdots & \ddots & \vdots & \vdots & \ddots & \vdots \\
                  0 & \ldots & 0 & d_{\frac{K}{2},\frac{K}{2}+1} & \cdots & d_{\frac{K}{2},K} \\
                  d_{\frac{K}{2}+1,1} & \cdots & d_{\frac{K}{2}+1,\frac{K}{2}} & 0 & \cdots & 0 \\
                  \vdots & \ddots & \vdots & \vdots & \ddots & \vdots \\
                  d_{K,1} & \cdots & d_{K,\frac{K}{2}} & 0 & \cdots & 0
                \end{array}
\right].
\end{equation}

Here, we separate the analysis into two cases.

\textbf{Case 1: $M$ is divisible by $\frac{K}{2}$.} Denote
\begin{equation}\label{desired_X}
\textbf{s}_{\oplus}=[{\bf s}_{1,2}^T+{\bf s}_{2,1}^T, \cdots, {\bf s}_{i,j}^T+{\bf s}_{j,i}^T, \cdots, {\bf s}_{K-1,K}^T+{\bf s}_{K,K-1}^T]^T
\end{equation}
as the network-coded symbol vector expected to obtain at the relay, where each ${\bf s}_{i,j}$ is a $d_{i,j} \times 1$ vector.

Let the $\frac{KM}{2} \times N$ compression matrix $\bf P$ be stacked by $\frac{2M}{K} \times N$ submatrices ${\bf P}_{i,j}$ by row. We design the compression matrix $\bf P$ by the method in the proof of \textit{Theorem 7}. Each submatrix ${\bf P}_{i,j}$ is designed as \eqref{X_null} at the top of the next page.
\begin{figure*}[ht]
\begin{equation}\label{X_null}
{\bf P}_{i,j}^T \subseteq \textbf{Null}~\big[{\bf H}_{1,r}~\cdots~{\bf H}_{i-1,r}~{\bf H}_{i+1,r}~\cdots~{\bf H}_{j-1,r}~{\bf H}_{j,r}~\cdots~{\bf H}_{K,r}\big]^T.
\end{equation}
\hrule
\end{figure*}
Then we design the precoding matrices ${\bf V}_{i,j}$ for each source node. Each pair of $M \times \frac{2M}{K}$ precoding matrices is designed as
\begin{align}\label{GSA_v_X}
\left[\begin{array}{c}
        {\bf V}_{i,j} \\
        {\bf V}_{j,i}
      \end{array}
\right] \subseteq \textbf{Null}~\left[{\bf P}{\bf H}_{i,r}~-{\bf P}{\bf H}_{j,r}\right].
\end{align}
Plugging \eqref{X_null} and \eqref{GSA_v_X} into \eqref{y_r_projection}, we can obtain the signals after compression as
\begin{align}\label{y_r_projection_X}
\hat{{\bf y}}_r={\bm \alpha} \textbf{s}_{\oplus} + {\bf P}{\bf n}_r
\end{align}
where ${\bm \alpha}$ is a diagonal matrix. Then, the network-coded symbol vector can be readily estimated from \eqref{y_r_projection_X}.

During the BC phase, we use the method of interference nulling to design the precoding matrix {\bf U}. We can write {\bf U} as follows.
\begin{equation}\label{UU_GXX}
{\bf U}=
\left[
\begin{array}{ccc}
 {\bf U}_{1}~{\bf U}_{2}~\cdots~{\bf U}_{\frac{K^2}{4}}
\end{array}
\right]
\end{equation}
where each ${\bf U}_i$ is an $N \times \frac{2M}{K}$ matrix and
\begin{align}\nonumber\label{UU3_GXX}
{\bf U}_1&\subseteq \textbf{Null}~\big[{\bf G}_{r,2}^T~{\bf G}_{r,3}^T~\cdots~{\bf G}_{r,\frac{K}{2}}^T~{\bf G}_{r,\frac{K}{2}+2}^T~\cdots~{\bf G}_{r,K}^T\big]^T~\\\nonumber
{\bf U}_2&\subseteq \textbf{Null}~\big[{\bf G}_{r,2}^T~{\bf G}_{r,3}^T~\cdots~{\bf G}_{r,\frac{K}{2}+1}^T~{\bf G}_{r,\frac{K}{2}+3}^T~\cdots~{\bf G}_{r,K}^T\big]^T~\\\nonumber
\vdots\\\nonumber
\vdots\\
{\bf U}_{\frac{K^2}{4}}&\subseteq \textbf{Null}~\big[{\bf G}_{r,1}^T~{\bf G}_{r,2}^T~\cdots~{\bf G}_{r,\frac{K}{2}-1}^T~{\bf G}_{r,\frac{K}{2}+1}^T~\cdots~{\bf G}_{r,K-1}^T\big]^T
\end{align}

We can see that ${\bf U}_i$ ($N \times \frac{2M}{K}$) exists if and only if $N-(K-2)M \geq \frac{2M}{K}$, or equivalently $N \geq \frac{(K^2-2K+2)M}{K}$. Hence, we can apply GSA-based transmission scheme when $M$ is a multiple of $\frac{K}{2}$ to achieve the DoF upper bound $KM$.

\textbf{Case 2: $M$ is not divisible by $\frac{K}{2}$.} We can utilize the ($\frac{K}{2}$)-symbol extension. The proof is similar to that in the previous subsection. We omit the detail proof here.

\subsection{Multi-user MIMO two-way relay channel}
The previous three subsections are for three special channel models. In this subsection, we consider the general multi-user MIMO two-way relay channel where the data switch matrix $\bf D$ can be arbitrary.

\textit{Theorem 8}: For the multi-user MIMO two-way relay channel, the total DoF of $\sum\limits_{i=1}^{K} \sum\limits_{j \in {\cal S}_i} {d_{i,j}}$ for any given data switch matrix $\bf D$ is achievable when $M \geq  \max_{i}\{\sum_{j \in {\cal S}_i} {d_{i,j}}\}$, and $N\geq (K-2)M+\max\{d_{i,j}\}$.
\begin{proof}
Let the $\frac{d_{total}}{2} \times N$ compression matrix $\bf P$ be stacked by submatrices ${\bf P}_{i,j}$ by row, where ${\bf P}_{i,j}$ is a $d_{i,j} \times N$ matrix. Note that ${\bf P}_{i,j}$ exists if and only if $\left[{\bf D}\right]_{i,j} \neq 0$. Construct each submatrix ${\bf P}_{i,j}$ as \eqref{null} at the top of the next page.
\begin{figure*}[ht]
\begin{equation}\label{null}
{\bf P}_{i,j}^T \subseteq \textbf{Null}~\big[{\bf H}_{1,r}~\cdots~{\bf H}_{i-1,r}~{\bf H}_{i+1,r}~\cdots~{\bf H}_{j-1,r}~{\bf H}_{j+1,r}~\cdots~{\bf H}_{K,r}\big]^T.
\end{equation}
\hrule
\end{figure*}
From \eqref{null}, we can obtain that ${\bf P}_{i,j}$ exists when
\begin{equation}\label{antenna_constraint}
N-(K-2)M \geq d_{i,j}.
\end{equation}

Remove the submatrices set $\{{\bf P}_{s,t} \mid s=i~\textrm{or}~s=j~\textrm{or}~t=i~\textrm{or}~t=j\}$ from $\bf P$, the remaining submatrix of $\bf P$ is defined as ${\bf F}_{i,j}$. From \eqref{null}, we can obtain that
\begin{equation}\label{F}
{\bf F}_{i,j}^T \subseteq \textbf{Null}~ \left[{\bf H}_{i,r}~-{\bf H}_{j,r}\right]^T
\end{equation}
The number of rows of the matrix ${\bf F}_{i,j}$ can be expressed as \eqref{rows} at the top of the next page.
\begin{figure*}[ht]
\begin{align}\nonumber
N_{i,j}&=\frac{d_{total}}{2}-(d_{1,i}+d_{2,i}+\cdots+d_{i-1,i}+d_{i,i+1}+\cdots+d_{i,K})\\\nonumber
&~~~-(d_{1,j}+d_{2,j}+\cdots+d_{j-1,j}+d_{j,j+1}+\cdots+d_{j,K})+d_{i,j}\\\nonumber
&=\frac{d_{total}}{2}-d_i-d_j+d_{i,j}\\\label{rows}
&\geq \frac{d_{total}}{2}-2M+d_{i,j}.
\end{align}
\hrule
\end{figure*}

Therefore, when $N \geq (K-2)M+\max\{d_{i,j}\}$, the compression matrix $\bf P$ can be constructed with the method of \eqref{null} and at least $\frac{d_{total}}{2}-2M+d_{i,j}$ row vectors of $\bf P$ will lie in the left null space of $\left[{\bf H}_{i,r}~-{\bf H}_{j,r}\right]$, for any $\left[{\bf D}\right]_{i,j} \neq 0$, which meets \textit{Theorem 4}.
\end{proof}

From \textit{Theorem 8}, we can obtain that the total DoF upper bound $KM$ is achievable for the multi-user MIMO two-way relay channel when $ \sum\limits_{j \in {\cal S}_i} {d_{i,j}}=M$, for all $i$'s, and $N\geq (K-2)M+\max\{d_{i,j}\}$.

\textit{Corollary 4}: The DoF upper bound of $KM$ is achievable for the $L$-cluster $K{'}=\frac{K}{L}$-user MIMO multiway relay channel when $\frac{N}{M} \geq \frac{(K{'}-1)(K-2)+1}{K{'}-1}$.
\begin{proof}
The proof follows directly from \textit{Theorem 8}. Note that $\max\{d_{i,j}\}$ is $\frac{M}{K{'}-1}$. Clearly, the region of antenna configuration for the DoF upper bound to be tight enlarges the one $N \geq LK{'}M=KM$ in \cite{Tian}.
\end{proof}

\section{Conclusion}
In this paper, we have introduced generalized signal alignment to analyze the achievable DoF for the general multi-user MIMO two-way relay channels, where each source node can exchange independent messages with an arbitrary set of other source nodes via the relay node. The proposed GSA is to align the signals to be exchanged between each source node pair at a compressed subspace of the relay. We provided the necessary and sufficient condition about the relay compression matrix for the GSA equation to hold. Using the proposed GSA, we have revealed new antenna configurations for achieving the maximum DoF of several special cases of the considered channel model, including the $K$-user MIMO Y channel, the multi-pair MIMO two-way relay channel, the generalized MIMO two-way X relay channnel, and the $L$-cluster MIMO multiway relay channel. We conclude that the proposed GSA represents a new and effective transmission framework towards the DoF analysis of a type of interference-limited wireless networks.

\section*{Appendix A: Proof of Theorem 1}
Consider that each source node can decode the $K-1$ intended messages with its own $K-1$ messages as side information. Then, if a genie provides that side information to the relay, the relay is able to decode the messages desired at that source node and the sum rate will not decrease.

\begin{figure}[t]
\begin{centering}
\includegraphics[scale=0.38]{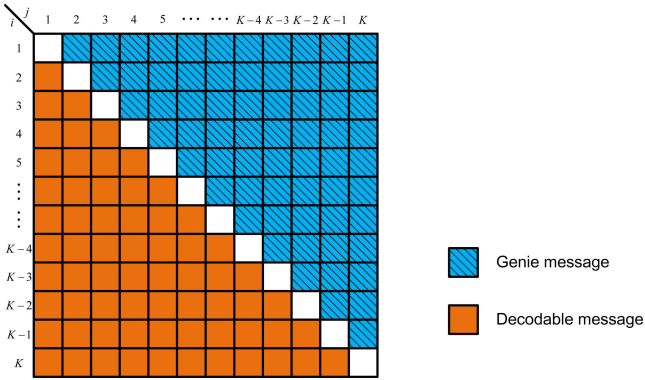}
\vspace{-0.1cm}
 \caption{Illustration for the genie information and the decodable messages at the relay for the $K$-user MIMO Y channel when $\frac{N}{M} \in \left(0, \frac{2K^2-2K}{K^2-K+2}\right]$}\label{Fig_genie_Y_1}
\end{centering}
\vspace{-0.2cm}
\end{figure}

We first consider the case when $\frac{N}{M} \in \left(0, \frac{2K^2-2K}{K^2-K+2}\right]$. As illustrated in Fig. \ref{Fig_genie_Y_1}, for each source node $i$, with $1 \leq i < K$, we provide the genie information $\{W_{i,j} \mid i+1 \leq j \leq K\}$ to the relay. Thus, the total genie information at the relay is ${\cal G}_1=\{W_{i,j} \mid i=1,2,\cdots,K-1; j=i+1,i+2,\cdots,K\}$ shown as upper triangle in Fig. \ref{Fig_genie_Y_1}. We can obtain the total transmission rate from nodes $\{i+1,i+2,\cdots,K\}$ to node $i$ during $n$ time slots as \eqref{genie_Y_1} at the top of the next page,
\begin{figure*}[ht]
\begin{subequations}\label{genie_Y_1}
\begin{align}\label{genie_Y_1_a}
&n(R_{i+1,i}+R_{i+2,i}+\cdots+R_{K,i})\\\label{genie_Y_1_b}
\leq & I(W_{i+1,i},W_{i+2,i},\cdots,W_{K,i};Y_i^n \mid W_{i,1}, W_{i,2}, \cdots, W_{i,i-1}, W_{i,i+1},\cdots, W_{i,K})+\epsilon(n)\\\label{genie_Y_1_c}
\leq & I(W_{i+1,i},W_{i+2,i},\cdots,W_{K,i};Y_r^n \mid W_{i,1}, W_{i,2}, \cdots, W_{i,i-1}, W_{i,i+1},\cdots, W_{i,K})+\epsilon(n)\\\label{genie_Y_1_d}
\leq & I(W_{i+1,i},W_{i+2,i},\cdots,W_{K,i};Y_r^n, {\cal G}_1\mid W_{i,1}, W_{i,2}, \cdots, W_{i,i-1}, W_{i,i+1},\cdots, W_{i,K})+\epsilon(n)\\\nonumber
= & I(W_{i+1,i},W_{i+2,i},\cdots,W_{K,i};{\cal G}_1\mid W_{i,1}, W_{i,2}, \cdots, W_{i,i-1}, W_{i,i+1},\cdots, W_{i,K})\\\label{genie_Y_1_e}
&~+I(W_{i+1,i},W_{i+2,i},\cdots,W_{K,i};Y_r^n \mid {\cal G}_1, W_{i,1}, W_{i,2}, \cdots, W_{i,i-1}, W_{i,i+1},\cdots, W_{i,K})+\epsilon(n)\\\label{genie_Y_1_f}
= & I(W_{i+1,i},W_{i+2,i},\cdots,W_{K,i};Y_r^n \mid {\cal G}_1, W_{i,1}, W_{i,2}, \cdots, W_{i,i-1}, W_{i,i+1},\cdots, W_{i,K})+\epsilon(n)\\\label{genie_Y_1_g}
= & I(W_{i+1,i},W_{i+2,i},\cdots,W_{K,i};Y_r^n \mid {\cal G}_1, W_{i,1}, W_{i,2}, \cdots, W_{i,i-1})+\epsilon(n).
\end{align}
\end{subequations}
\hrule
\end{figure*}
where $\epsilon(n)$ represents that $\lim\limits_{n \rightarrow \infty} \frac{\epsilon(n)}{n}=0$ and $X_i$ represents all the messages transmitted from source node $i$. Here, \eqref{genie_Y_1_b} follows from the Fano's inequality,  \eqref{genie_Y_1_c} is obtained via the data processing inequality because $Y_r-X_r-Y_i$ forms a Markov chain,  \eqref{genie_Y_1_d} is obtained because adding genie signals does not reduce the capacity region, and \eqref{genie_Y_1_f} follows from the fact that the first term in \eqref{genie_Y_1_e} is zero.

Adding \eqref{genie_Y_1} from $i=1$ to $K-1$, we can obtain \eqref{genie_Y_1_sum} at the top of the next page.
\begin{figure*}[ht]
\begin{subequations}\label{genie_Y_1_sum}
\begin{align}
&n\left(\sum\limits_{i=1}^{K-1} \sum\limits_{j=i+1}^{K} R_{j,i}\right)\\
\leq & I(\{W_{j,i} \mid i=1,2,\cdots,K-1; j=i+1,i+2,\cdots,K\};Y_r^n \mid {\cal G}_1)+\epsilon(n)\\
\leq & h(Y_r^n \mid {\cal G}_1)+\epsilon(n)\\
\leq & nN\textrm{log}P+\epsilon(n)
\end{align}
\end{subequations}
\hrule
\end{figure*}
Dividing $n\textrm{log}P$ to both sides of \eqref{genie_Y_1_sum} and letting $n\rightarrow \infty$ and $P\rightarrow \infty$, we can obtain the total DoF upper bound as
\begin{align}
d_{total}=\sum\limits_{i=1}^{K} \sum\limits_{j=1}^K d_{i,j} \leq 2N.
\end{align}

\begin{figure}[t]
\begin{centering}
\includegraphics[scale=0.38]{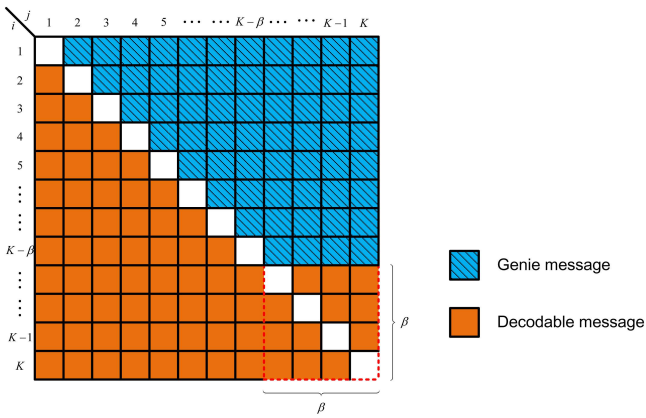}
\vspace{-0.1cm}
 \caption{Illustration for the genie information and the decodable messages at the relay for the $K$-user MIMO Y channel when $\frac{N}{M} \in \left(\beta, \frac{(\beta+1)(K(K-1)+(\beta)(\beta-1))}{K(K-1)+(\beta+1)\beta}\right]$}\label{Fig_genie_Y_beta}
\end{centering}
\vspace{-0.2cm}
\end{figure}

Next, we consider the case when $\frac{N}{M} \in \left(\beta, \frac{(\beta+1)(K(K-1)+\beta(\beta-1))}{K(K-1)+(\beta+1)\beta}\right]$ for each $\beta \in \{2,3,4,\cdots,K-2\}$. As illustrated in Fig. \ref{Fig_genie_Y_beta}, for each source node $i$, with $1 \leq i \leq K-\beta$, we provide the genie information $\{W_{i,j} \mid i+1 \leq j \leq K\}$ to the relay. Thus, the total genie information at the relay is ${\cal G}_2=\{W_{i,j} \mid i=1,2,\cdots,K-\beta; j=i+1,i+2,\cdots,K\}$. Similar to \eqref{genie_Y_1}, we can obtain \eqref{genie_Y_beta} at the top of the next page,
\begin{figure*}[ht]
\begin{align}\label{genie_Y_beta}
n(R_{i+1,i}+R_{i+2,i}+\cdots+R_{K,i})\leq I(W_{i+1,i},W_{i+2,i},\cdots,W_{K,i};Y_r^n \mid {\cal G}_2, W_{i,1}, W_{i,2}, \cdots, W_{i,i-1})+\epsilon(n)
\end{align}
\hrule
\end{figure*}
for $i=1,2,\cdots,K-\beta$. Compared with Fig. \ref{Fig_genie_Y_1}, the difference is that no genie information is provided for the source nodes $\{K-\beta+1, K-\beta+2,\cdots,K\}$. Adding \eqref{genie_Y_beta} from $i=1$ to $K-\beta$, we can obtain \eqref{genie_Y_beta_sum} at the top of the next page
\begin{figure*}[ht]
\begin{subequations}\label{genie_Y_beta_sum}
\begin{align}
&n\left(\sum\limits_{i=1}^{K-\beta} \sum\limits_{j=i+1}^{K} R_{j,i}\right)\\
\leq & I(\{W_{j,i} \mid i=1,2,\cdots,K-\beta; j=i+1,i+2,\cdots,K\};Y_r^n \mid {\cal G}_2)+\epsilon(n)\\
= & h(Y_r^n \mid {\cal G}_2)-h(Y_r^n \mid {\cal G}_2,\{W_{j,i} \mid i=1,2,\cdots,K-\beta; j=i+1,i+2,\cdots,K\})+\epsilon(n)\\
= & h(Y_r^n \mid {\cal G}_2)-h(Y_r^n \mid X_1^n,X_2^n,\cdots,X_{K-\beta}^n,\{W_{j,i} \mid i\in[1,K-\beta]; j\in [K-\beta+1,K]\})+\epsilon(n)\\
= & h(Y_r^n \mid {\cal G}_2)-h(X_{K-\beta+1}^n,X_{K-\beta+2}^n,\cdots,X_{K}^n \mid \{W_{j,i} \mid i\in[1,K-\beta]; j\in [K-\beta+1,K]\})+n\epsilon(\textrm{log}P)+\epsilon(n)\\
= & h(Y_r^n \mid {\cal G}_2)-H(\{W_{i,j} \mid i \in [K-\beta+1,K];j \in [K-\beta+1,K];j \neq i\})+n\epsilon(\textrm{log}P)+\epsilon(n)\\
\leq & nN\textrm{log}P-n(\{R_{i,j} \mid i \in [K-\beta+1,K]; j \in [K-\beta+1,K]; j \neq i\})+n\epsilon(\textrm{log}P)+\epsilon(n)
\end{align}
\end{subequations}
\hrule
\end{figure*}
and we have \eqref{R_Y} at the top of the next page
\begin{figure*}[ht]
\begin{align}\label{R_Y}
n\left(\sum\limits_{i=1}^{K-1} \sum\limits_{j=i+1}^{K} R_{j,i}+\sum\limits_{i=K-\beta+1}^{K-1} \sum\limits_{j=i+1}^{K} R_{i,j}\right)\leq nN\textrm{log}P+n\epsilon(\textrm{log}P)+\epsilon(n).
\end{align}
\hrule
\end{figure*}
We can obtain similar equations to \eqref{R_Y} by replacing the $\beta$ source nodes $\{K-\beta+1,K-\beta+2,\cdots,K\}$ to any other $\beta$ source nodes. Then dividing $n\textrm{log}P$ to both sides of \eqref{R_Y} and letting $n\rightarrow \infty$ and $P\rightarrow \infty$, we can obtain the total DoF upper bound as
\begin{align}
d_{total}=\sum\limits_{i=1}^{K} \sum\limits_{j=1}^K d_{i,j} \leq \frac{2K(K-1)N}{K(K-1)+\beta(\beta-1)}.
\end{align}

Third, we consider the case when $\frac{N}{M} \in \left(\frac{\beta(K(K-1)+(\beta-1)(\beta-2))}{K(K-1)+\beta (\beta-1)}, \beta\right]$. We prove this by contradiction. If $\frac{N}{M}  \in \left(\frac{\beta(K(K-1)+(\beta-1)(\beta-2))}{K(K-1)+\beta (\beta-1)}, \beta\right]$ and $d_{total}^a > \frac{2\beta K(K-1)M}{K(K-1)+\beta (\beta-1)}$, where $d_{total}^a$ represents the achivable total DoF, then we increase $N$ to $N_1$ such that $\frac{N_1}{M} =\beta$. Utilizing the antenna deactivation, $d_{total}^a > \frac{2\beta K(K-1)M}{K(K-1)+\beta (\beta-1)}=\frac{2 K(K-1)N_1}{K(K-1)+\beta (\beta-1)}$ can be achieved. However, this contradicts with that the DoF upper bound is $\frac{2 K(K-1)N_1}{K(K-1)+\beta (\beta-1)}$ when $\frac{N_1}{M} =\beta$. Hence, the DoF upper bound of the case when $\frac{N}{M} \in \left(\frac{\beta(K(K-1)+(\beta-1)(\beta-2))}{K(K-1)+\beta (\beta-1)}, \beta\right]$ is $\frac{2\beta K(K-1)M}{K(K-1)+\beta (\beta-1)}$.

Finally, we consider the case when $\frac{N}{M} \in \left(\frac{K^2-3K+3}{K-1}, +\infty \right)$. In this case, we notice that the DoF per user could not be larger than $M$. Thus, $KM$ is the DoF upper bound for this case.

\section*{Appendix B: Proof of Theorem 2}
\begin{figure}[t]
\begin{centering}
\includegraphics[scale=0.38]{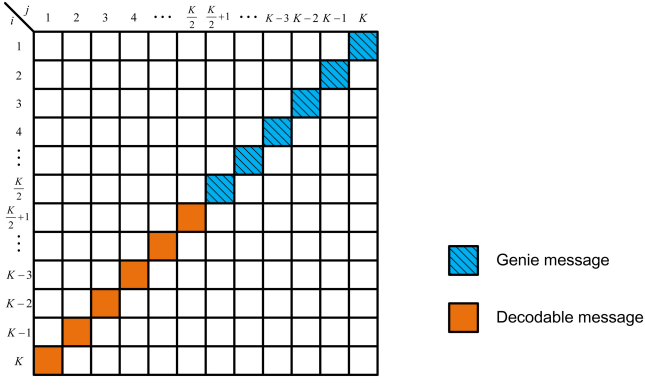}
\vspace{-0.1cm}
 \caption{Illustration for the genie information and the decodable messages for the multi-pair MIMO two-way relay channel at the relay when $\frac{N}{M} \in \left(0, \frac{2K}{K+2}\right]$}\label{Fig_genie_MP_1}
\end{centering}
\vspace{-0.2cm}
\end{figure}

The idea of this proof is similar to \textit{Theorem 1}. We first consider the case when $\frac{N}{M} \in \left(0, \frac{2K}{K+2}\right]$. As illustrated in Fig. \ref{Fig_genie_MP_1}, we provide the genie information ${\cal G}_1=\{W_{i,j} \mid i=1,2,\cdots,\frac{K}{2}; j=K+1-i\}$ to the relay. We can obtain the total transmission rate from nodes $K+1-i$ to node $i$ during $n$ time slots as
\begin{subequations}\label{genie_MP_1}
\begin{align}
&n(R_{K+1-i,i})\\
\leq & I(W_{K+1-i,i};Y_i^n \mid W_{i,K+1-i})+\epsilon(n)\\
\leq & I(W_{K+1-i,i};Y_r^n \mid W_{i,K+1-i})+\epsilon(n)\\
\leq & I(W_{K+1-i,i};Y_r^n, {\cal G}_1 \mid W_{i,K+1-i})+\epsilon(n)\\\nonumber
= & I(W_{K+1-i,i};{\cal G}_1 \mid W_{i,K+1-i})\\
&+I(W_{K+1-i,i};Y_r^n \mid {\cal G}_1, W_{i,K+1-i})+\epsilon(n)\\
= & I(W_{K+1-i,i};Y_r^n \mid {\cal G}_1, W_{i,K+1-i})+\epsilon(n)\\
= & I(W_{K+1-i,i};Y_r^n \mid {\cal G}_1)+\epsilon(n)
\end{align}
\end{subequations}
Adding \eqref{genie_MP_1} from $i=1$ to $\frac{K}{2}$, we can obtain
\begin{subequations}\label{genie_MP_1_sum}
\begin{align}
&n\left(\sum\limits_{i=1}^{\frac{K}{2}} R_{K+1-i,i}\right)\\
\leq & I(\{W_{j,i} \mid i=1,2,\cdots,\frac{K}{2}; j=K+1-i\};Y_r^n \mid {\cal G}_1)+\epsilon(n)\\
\leq & h(Y_r^n \mid {\cal G}_1)+\epsilon(n)\\
\leq & nN\textrm{log}P+\epsilon(n)
\end{align}
\end{subequations}
Dividing $n\textrm{log}P$ to both sides of \eqref{genie_MP_1_sum} and letting $n\rightarrow \infty$ and $P\rightarrow \infty$, we obtain the total DoF upper bound as
\begin{align}
d_{total}=\sum\limits_{i=1}^{K} d_{i,K+1-i} \leq 2N.
\end{align}

\begin{figure}[t]
\begin{centering}
\includegraphics[scale=0.38]{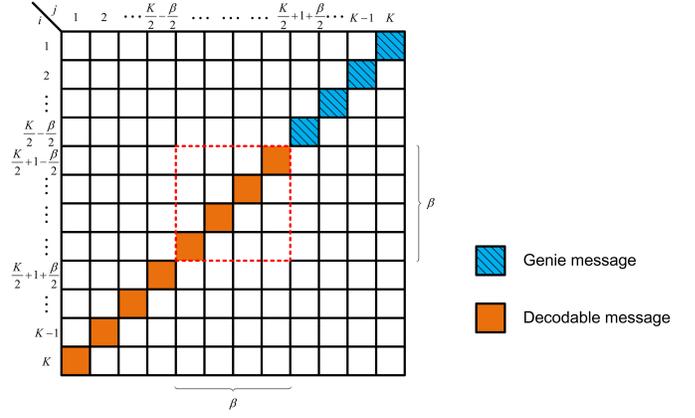}
\vspace{-0.1cm}
 \caption{Illustration for the genie information and the decodable messages at the relay for the multi-pair MIMO two-way relay channel when $\frac{N}{M} \in \left(\beta, \frac{(\beta+1)(K(K-1)+(\beta)(\beta-1))}{K(K-1)+(\beta+1)\beta}\right]$}\label{Fig_genie_MP_beta}
\end{centering}
\vspace{-0.2cm}
\end{figure}

Next, we consider the case when $\frac{N}{M} \in \left(\beta, \frac{(\beta+2)(K+\beta)}{K+\beta+2}\right]$, for each $\beta \in \{2,4,\cdots,K-2\}$. Note that $\beta$ is even here. As illustrated in Fig. \ref{Fig_genie_MP_beta}, for each source node $i$, with $1 \leq i \leq \frac{K}{2}-\frac{\beta}{2}$, we provide the genie information $\{W_{i,j} \mid j=K+1-i\}$ to the relay. Thus, the total genie information at the relay is ${\cal G}_2=\{W_{i,j} \mid i=1,2,\cdots,\frac{K}{2}-\frac{\beta}{2}; j=K+1-i\}$ to the relay. Similar to \eqref{genie_MP_1}, we can obtain
\begin{align}\label{genie_MP_beta}
n(R_{K+1-i,i})\leq I(W_{K+1-i,i};Y_r^n \mid {\cal G}_2)+\epsilon(n)
\end{align}
for $i=1,2,\cdots,K-\beta$. Compared with Fig. \ref{Fig_genie_MP_1}, the difference is that no genie information is provided for the source nodes $\left\{\frac{K}{2}-\frac{\beta}{2}+1, \frac{K}{2}-\frac{\beta}{2}+2,\cdots,\frac{K}{2}\right\}$. Adding \eqref{genie_MP_beta} from $i=1$ to $\frac{K}{2}-\frac{\beta}{2}$, we can obtain \eqref{genie_MP_beta_sum} at the top of the next page.
\begin{figure*}[ht]
\begin{subequations}\label{genie_MP_beta_sum}
\begin{align}
&n\left(\sum\limits_{i=1}^{\frac{K}{2}-\frac{\beta}{2}} R_{K+1-i,i}\right)\\
\leq & I\left(\left\{W_{K+1-i,i} \mid i=1,2,\cdots,\frac{K}{2}-\frac{\beta}{2}\right\};Y_r^n \mid {\cal G}_2\right)+\epsilon(n)\\
= & h(Y_r^n \mid {\cal G}_2)-h\left(Y_r^n \mid {\cal G}_2, \left\{W_{K+1-i,i} \mid i=1,2,\cdots,\frac{K}{2}-\frac{\beta}{2}\right\}\right)+\epsilon(n)\\\nonumber
= & h(Y_r^n \mid {\cal G}_2)-h\left(Y_r^n \mid X_1^n,\cdots,X_{\frac{K}{2}-\frac{\beta}{2}}^n,X_{\frac{K}{2}+\frac{\beta}{2}+1}^n \cdots, X_{K}^n, \left\{W_{K+1-i,i} \mid i\in\left[1,\frac{K}{2}-\frac{\beta}{2}\right]\right\}\right)+\epsilon(n)\\
= & h(Y_r^n \mid {\cal G}_2)-h(X_{\frac{K}{2}-\frac{\beta}{2}+1}^n,X_{\frac{K}{2}-\frac{\beta}{2}+2}^n,\cdots,X_{\frac{K}{2}+\frac{\beta}{2}}^n)
+n\epsilon(\textrm{log}P)+\epsilon(n)\\
= & h(Y_r^n \mid {\cal G}_2)-H\left(\left\{W_{i,K+1-i} \mid i \in \left[\frac{K}{2}-\frac{\beta}{2}+1, \frac{K}{2}+\frac{\beta}{2}\right] \right\}\right)+n\epsilon(\textrm{log}P)+\epsilon(n)\\
\leq & nN\textrm{log}P-n\left(\left\{R_{i,K+1-i} \mid i \in \left[\frac{K}{2}-\frac{\beta}{2}+1, \frac{K}{2}+\frac{\beta}{2}\right]\right\}\right)+n\epsilon(\textrm{log}P)+\epsilon(n).
\end{align}
\end{subequations}
\hrule
\end{figure*}
Then we have
\begin{align}\label{R_MP}
n\left(\sum\limits_{i=1}^{\frac{K}{2}+\frac{\beta}{2}} R_{K+1-i,i}\right)\leq  nN\textrm{log}P+n\epsilon(\textrm{log}P)+\epsilon(n).
\end{align}
We can obtain similar equations to \eqref{R_MP} by replacing the $\beta$ source nodes $\left\{\frac{K}{2}-\frac{\beta}{2}+1,\frac{K}{2}-\frac{\beta}{2}+2,\cdots,\frac{K}{2}+\frac{\beta}{2}\right\}$ to any other $\beta$ source nodes. Then dividing $n\textrm{log}P$ to both sides of \eqref{R_MP} and letting $n\rightarrow \infty$ and $P\rightarrow \infty$, we can obtain the total DoF upper bound as
\begin{align}
d_{total}=\sum\limits_{i=1}^{K} d_{i,K+1-i} \leq \frac{2KN}{K+\beta}.
\end{align}

Thirdly, we consider the case when $\frac{N}{M} \in \left(\frac{\beta(K+\beta-2)}{K+\beta}, \beta\right]$. We prove this by contradiction. If $\frac{N}{M} \in \left(\frac{\beta(K+\beta-2)}{K+\beta}, \beta\right]$ and $d_{total}^a > \frac{2\beta KM}{K+\beta}$, where $d_{total}^a$ represents the achivable total DoF, then we increase $N$ to $N_1$ such that $\frac{N_1}{M} =\beta$. Utilizing the antenna deactivation, $d_{total}^a > \frac{2\beta KM}{K+\beta}=\frac{2KN_1}{K+\beta}$ can be achieved. However, this contradicts with that when $\frac{N_1}{M} =\beta$, the DoF upper bound is $\frac{2KN_1}{K+\beta}$. Hence, the DoF upper bound of the case when $\frac{N}{M} \in \left(\frac{\beta(K+\beta-2)}{K+\beta}, \beta\right]$ is $\frac{2\beta KM}{K+\beta}$.

Finally, we consider the case when $\frac{N}{M} \in \big(K-1, +\infty \big)$. In this case, we notice that the DoF per user could not be larger than $M$. Thus, $KM$ is the DoF upper bound for this case.

\section*{Appendix C: Proof of Theorem 3}
\begin{figure}[t]
\begin{centering}
\includegraphics[scale=0.38]{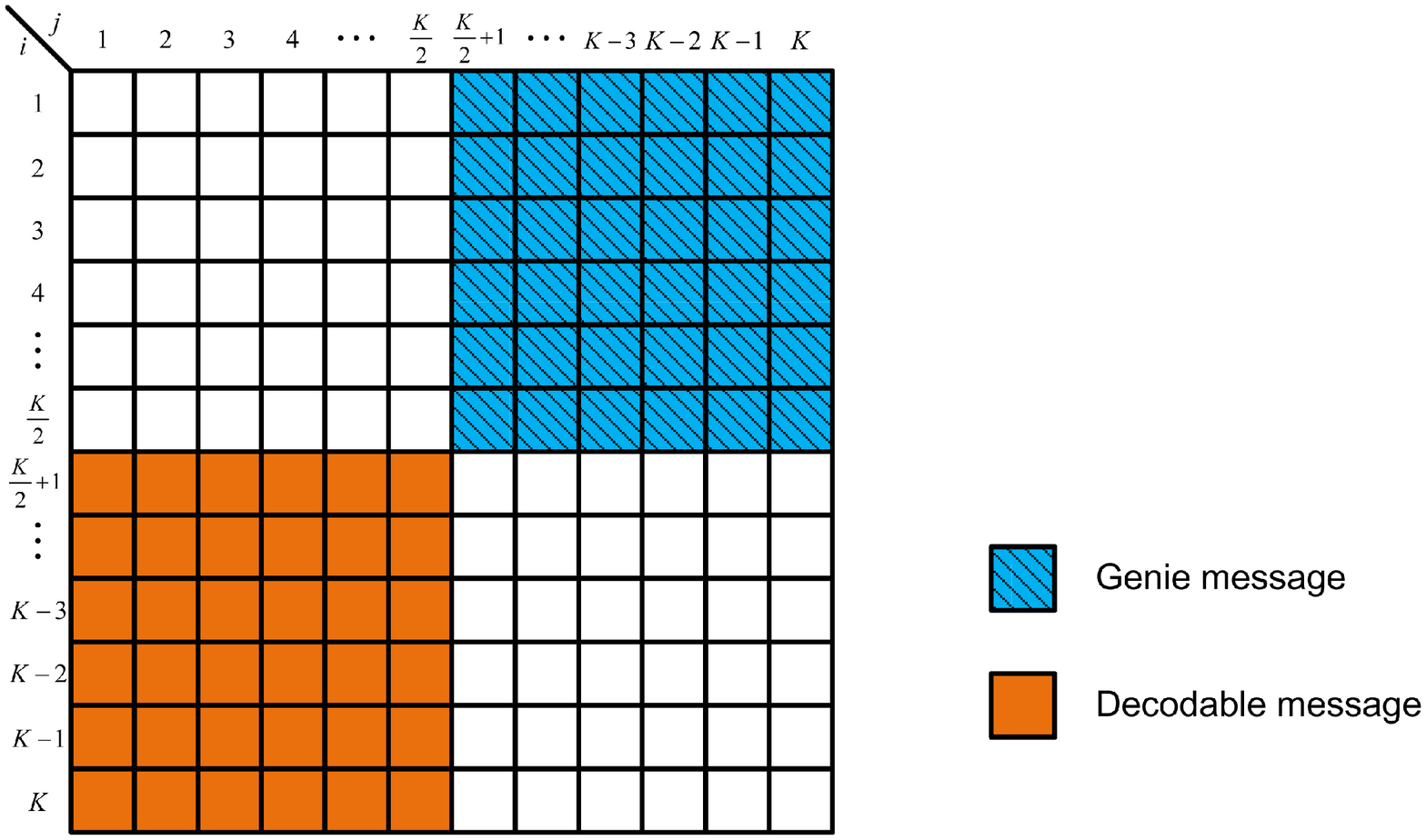}
\vspace{-0.1cm}
 \caption{Illustration for the genie information and the decodable messages at the relay for the generalized MIMO two-way X relay channel when $\frac{N}{M} \in \left(0, \frac{2K^2}{K^2+4}\right]$}\label{Fig_genie_X_1}
\end{centering}
\vspace{-0.2cm}
\end{figure}

The idea of this proof is similar to \textit{Theorem 1}. We first consider the case when $\frac{N}{M} \in \left(0, \frac{2K}{K+2}\right]$. As illustrated in Fig. \ref{Fig_genie_X_1}, for each source node $i$, with $1 \leq i \leq \frac{K}{2}$, we provide the genie information $\left\{W_{i,j} \mid \frac{K}{2}+1 \leq j \leq K\right\}$ to the relay. Thus, the total genie information at the relay is ${\cal G}_1=\left\{W_{i,j} \mid i=1,2,\cdots,\frac{K}{2}; j=\frac{K}{2}+1,\frac{K}{2}+2,\cdots,K\right\}$ to the relay. We can obtain the total transmission rate from nodes $\left\{\frac{K}{2}+1,\frac{K}{2}+2,\cdots,K\right\}$ to node $i$ during $n$ time slots as \eqref{genie_X_1} at the top of the next page.
\begin{figure*}[ht]
\begin{subequations}\label{genie_X_1}
\begin{align}
&n(R_{\frac{K}{2}+1,i}+R_{\frac{K}{2}+2,i}+\cdots+R_{K,i})\\
\leq & I(W_{\frac{K}{2}+1,i},W_{\frac{K}{2}+2,i},\cdots,W_{K,i};Y_i^n \mid W_{i,\frac{K}{2}+1}, W_{i,\frac{K}{2}+2}, \cdots, W_{i,K})+\epsilon(n)\\
\leq & I(W_{\frac{K}{2}+1,i},W_{\frac{K}{2}+2,i},\cdots,W_{K,i};Y_r^n \mid W_{i,\frac{K}{2}+1}, W_{i,\frac{K}{2}+2}, \cdots, W_{i,K})+\epsilon(n)\\
\leq & I(W_{\frac{K}{2}+1,i},W_{\frac{K}{2}+2,i},\cdots,W_{K,i};Y_r^n, {\cal G}_1 \mid W_{i,\frac{K}{2}+1}, W_{i,\frac{K}{2}+2}, \cdots, W_{i,K})+\epsilon(n)\\
= & I(W_{\frac{K}{2}+1,i},W_{\frac{K}{2}+2,i},\cdots,W_{K,i}; {\cal G}_1 \mid W_{i,\frac{K}{2}+1}, W_{i,\frac{K}{2}+2}, \cdots, W_{i,K})\\
&~+I(W_{\frac{K}{2}+1,i},W_{\frac{K}{2}+2,i},\cdots,W_{K,i};Y_r^n \mid {\cal G}_1, W_{i,\frac{K}{2}+1}, W_{i,\frac{K}{2}+2}, \cdots, W_{i,K})+\epsilon(n)\\
= & I(W_{\frac{K}{2}+1,i},W_{\frac{K}{2}+2,i},\cdots,W_{K,i};Y_r^n \mid {\cal G}_1, W_{i,\frac{K}{2}+1}, W_{i,\frac{K}{2}+2}, \cdots, W_{i,K})+\epsilon(n)\\
= & I(W_{i+1,i},W_{i+2,i},\cdots,W_{K,i};Y_r^n \mid {\cal G}_1)+\epsilon(n)
\end{align}
\end{subequations}
\hrule
\end{figure*}
Adding \eqref{genie_X_1} from $i=1$ to $\frac{K}{2}$, we can obtain \eqref{genie_X_1_sum} at the top of the next page.
\begin{figure*}[ht]
\begin{subequations}\label{genie_X_1_sum}
\begin{align}
&n\left(\sum\limits_{i=1}^{\frac{K}{2}} \sum\limits_{j=\frac{K}{2}+1}^{K} R_{j,i}\right)\\
\leq & I\left(\left\{W_{j,i} \mid i=1,2,\cdots,\frac{K}{2}; j=\frac{K}{2}+1,\frac{K}{2}+2,\cdots,K\right\};Y_r^n \mid {\cal G}_1\right)+\epsilon(n)\\
\leq & h(Y_r^n \mid {\cal G}_1)+\epsilon(n)\\
\leq & nN\textrm{log}P+\epsilon(n)
\end{align}
\end{subequations}
\hrule
\end{figure*}
Dividing $n\textrm{log}P$ to both sides of \eqref{genie_X_1_sum} and letting $n\rightarrow \infty$ and $P\rightarrow \infty$, we can obtain the total DoF upper bound as
\begin{align}
d_{total}=\sum\limits_{i=1}^{\frac{K}{2}} \sum\limits_{j \in {\cal S}_i} d_{i,j} \leq \frac{K^2}{2}\frac{4N}{K^2} = 2N.
\end{align}

\begin{figure}[t]
\begin{centering}
\includegraphics[scale=0.38]{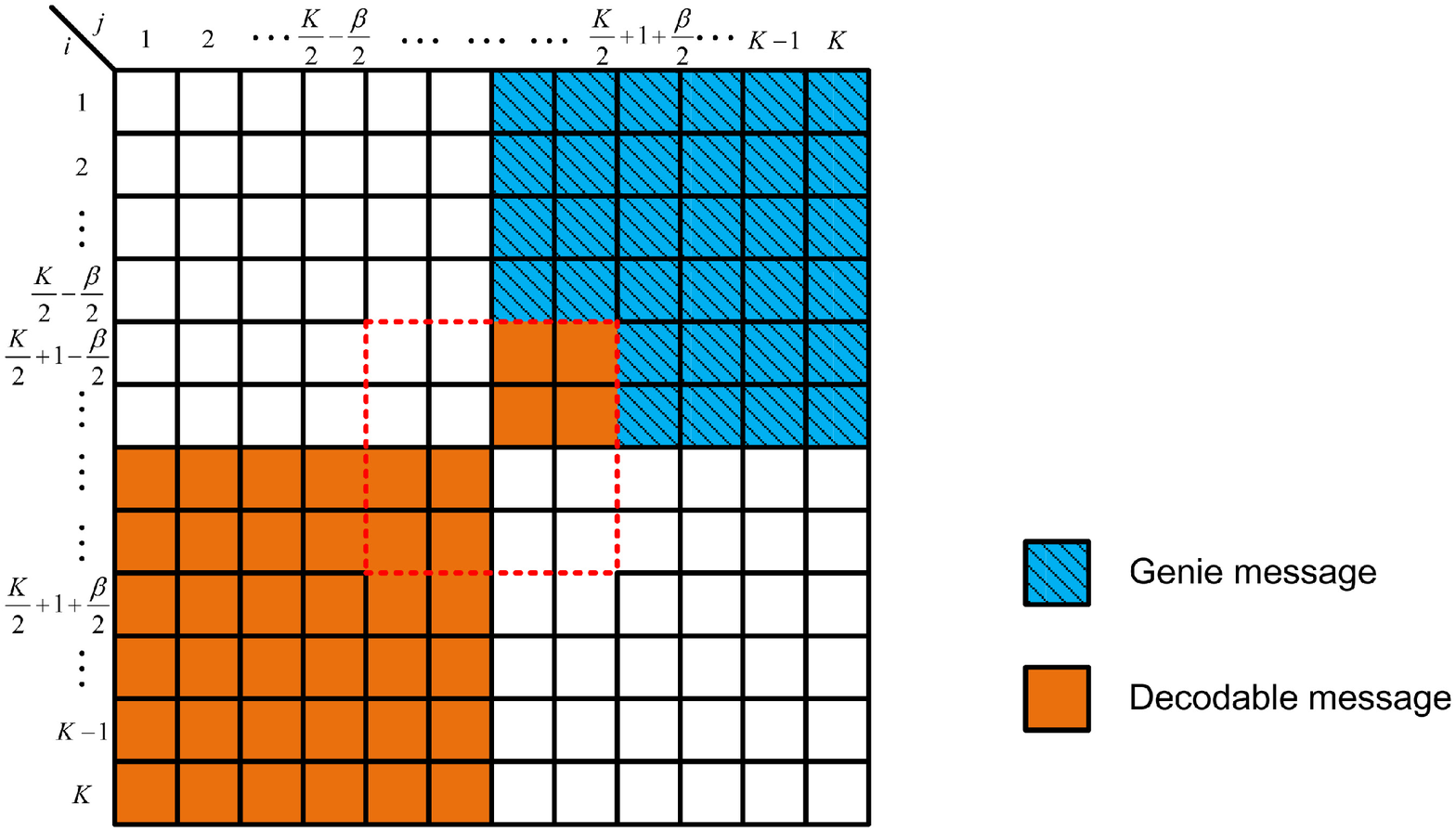}
\vspace{-0.1cm}
 \caption{Illustration for the genie information and the decodable messages at the relay for the generalized MIMO two-way X relay channel when $\frac{N}{M} \in \big(\beta, \frac{(K^2+\beta^2)(\beta+2)}{K^2+(\beta+2)^2}\big]$}\label{Fig_genie_X_beta}
\end{centering}
\vspace{-0.2cm}
\end{figure}

Next, we consider the case when $\frac{N}{M} \in \left(\beta, \frac{(K^2+\beta^2)(\beta+2)}{K^2+(\beta+2)^2}\right]$, for each $\beta \in \left\{2,4,\cdots,K-2\right\}$. As illustrated in Fig. \ref{Fig_genie_X_beta}, we provide the genie information ${\cal G}_2=\{W_{i,j} \mid i=1,2,\cdots,\frac{K}{2}-\frac{\beta}{2}; j=\frac{K}{2}+1,\frac{K}{2}+2,\cdots,K\}\cup \{W_{i,j} \mid i=\frac{K}{2}-\frac{\beta}{2}+1,\cdots,\frac{K}{2}; j=\frac{K}{2}+\frac{\beta}{2}+1,\cdots,K\}$ to the relay. Similar to \eqref{genie_X_1}, we can obtain
\begin{align}\nonumber
&n(R_{i+1,i}+R_{i+2,i}+\cdots+R_{K,i})\\\label{genie_X_beta_1}
\leq & I(W_{i+1,i},W_{i+2,i},\cdots,W_{K,i};Y_r^n \mid {\cal G}_2)+\epsilon(n)
\end{align}
where $i=1,2,\cdots,\frac{K}{2}-\frac{\beta}{2}$ and \eqref{genie_X_beta_2} for $i=\frac{K}{2}-\frac{\beta}{2}+1,\cdots,\frac{K}{2}$  at the top of the next page.
\begin{figure*}[ht]
\begin{align}\label{genie_X_beta_2}
n(R_{i+1,i}+R_{i+2,i}+\cdots+R_{K,i}) \leq I(W_{i+1,i},W_{i+2,i},\cdots,W_{K,i};Y_r^n \mid {\cal G}_2, W_{i,\frac{K}{2}+1}, W_{i,\frac{K}{2}+2}, \cdots, W_{i,\frac{K}{2}+\frac{\beta}{2}})+\epsilon(n)
\end{align}
\hrule
\end{figure*}

Adding \eqref{genie_X_beta_1} and \eqref{genie_X_beta_2} from $i=1$ to $\frac{K}{2}-\frac{\beta}{2}$, we can obtain \eqref{genie_X_beta_sum} at the top of the next page.
\begin{figure*}[ht]
\begin{subequations}\label{genie_X_beta_sum}
\begin{align}
&n\left(\sum\limits_{i=1}^{\frac{K}{2}} \sum\limits_{j \in {\cal S}_i} R_{j,i}\right)\\\nonumber
\leq & I\left(\left\{W_{i,j} \mid i=1,2,\cdots,\frac{K}{2}-\frac{\beta}{2}; j=\frac{K}{2}+1,\frac{K}{2}+2,\cdots,K\right\},\left\{W_{i,j} \mid i=\frac{K}{2}-\frac{\beta}{2}+1,\cdots,\frac{K}{2};\right.\right.\\
&\left.\left.~ j=\frac{K}{2}+\frac{\beta}{2}+1,\cdots,K\right\};Y_r^n \mid {\cal G}_2\right)+\epsilon(n)\\\nonumber
= & h(Y_r^n \mid {\cal G}_2)-h\left(Y_r^n \mid {\cal G}_2,\left\{W_{i,j} \mid i=1,2,\cdots,\frac{K}{2}-\frac{\beta}{2}; j=\frac{K}{2}+1,\frac{K}{2}+2,\cdots,K\right\},\right.\\
&\left.~\left\{W_{i,j} \mid i=\frac{K}{2}-\frac{\beta}{2}+1,\cdots,\frac{K}{2}; j=\frac{K}{2}+\frac{\beta}{2}+1,\cdots,K\right\}\right)+\epsilon(n)\\\nonumber
= & h(Y_r^n \mid {\cal G}_2)-h\left(Y_r^n \mid X_1^n,\cdots,X_{\frac{K}{2}-\frac{\beta}{2}}^n,X_{\frac{K}{2}+\frac{\beta}{2}+1}^n,\cdots,X_{K}^n,\left\{W_{j,i} \mid i\in\left[\frac{K}{2}-\frac{\beta}{2}+1,\frac{K}{2}\right]; \right.\right.\\
&\left.\left.~j\in \left[\frac{K}{2}+1,\frac{K}{2}+\frac{\beta}{2}\right]\right\},\left\{W_{j,i} \mid i\in\left[\frac{K}{2}+1,\frac{K}{2}+\frac{\beta}{2}\right]; j\in \left[\frac{K}{2}-\frac{\beta}{2}+1,\frac{K}{2}\right]\right\}\right)+\epsilon(n)\\\nonumber
= & h(Y_r^n \mid {\cal G}_2)-h\left(X_{\frac{K}{2}-\frac{\beta}{2}+1}^n,\cdots,X_{\frac{K}{2}+\frac{\beta}{2}}^n \mid \left\{W_{j,i} \mid i\in\left[\frac{K}{2}-\frac{\beta}{2}+1,\frac{K}{2}\right]; j\in \left[\frac{K}{2}+1,\frac{K}{2}+\frac{\beta}{2}\right]\right\},\right.\\\nonumber
&\left.~\left\{W_{j,i} \mid i\in\left[\frac{K}{2}+1,\frac{K}{2}+\frac{\beta}{2}\right]; j\in \left[\frac{K}{2}-\frac{\beta}{2}+1,\frac{K}{2}\right]\right\}\right)+n\epsilon(\textrm{log}P)+\epsilon(n)\\\nonumber
= & h(Y_r^n \mid {\cal G}_2)-H\left(\left\{W_{i,j} \mid i\in\left[\frac{K}{2}-\frac{\beta}{2}+1,\frac{K}{2}\right]; j\in \left[\frac{K}{2}+1,\frac{K}{2}+\frac{\beta}{2}\right]\right\},\right.\\
&\left.~\left\{W_{i,j} \mid i\in\left[\frac{K}{2}+1,\frac{K}{2}+\frac{\beta}{2}\right]; j\in \left[\frac{K}{2}-\frac{\beta}{2}+1,\frac{K}{2}\right]\right\}\right)+n\epsilon(\textrm{log}P)+\epsilon(n)\\\nonumber
\leq & nN\textrm{log}P-n\left(\left\{R_{i,j} \mid i\in\left[\frac{K}{2}-\frac{\beta}{2}+1,\frac{K}{2}\right]; j\in \left[\frac{K}{2}+1,\frac{K}{2}+\frac{\beta}{2}\right]\right\},\right.\\
&\left.~\left\{R_{i,j} \mid i\in[\frac{K}{2}+1,\frac{K}{2}+\frac{\beta}{2}]; j\in \left[\frac{K}{2}-\frac{\beta}{2}+1,\frac{K}{2}\right]\right\}\right)+n\epsilon(\textrm{log}P)+\epsilon(n)
\end{align}
\end{subequations}
\hrule
\end{figure*}
Then we have
\begin{align}\nonumber
&n\left(\sum\limits_{i=1}^{\frac{K}{2}} \sum\limits_{j=\frac{K}{2}+1}^{K} R_{j,i}+\sum\limits_{i=\frac{K}{2}-\frac{\beta}{2}+1}^{\frac{K}{2}} \sum\limits_{j=\frac{K}{2}}^{\frac{K}{2}+\frac{\beta}{2}+1} R_{i,j}\right)\\\label{R_X}
\leq & nN\textrm{log}P+n\epsilon(\textrm{log}P)+\epsilon(n).
\end{align}
We can obtain similar equations to \eqref{R_X} by replacing the $\beta$ source nodes $\left\{\frac{K}{2}-\frac{\beta}{2}+1,\frac{K}{2}-\frac{\beta}{2}+2,\cdots,\frac{K}{2}+\frac{\beta}{2}\right\}$ to any other $\beta$ source nodes. Then dividing $n\textrm{log}P$ to both sides of \eqref{R_X} and letting $n\rightarrow \infty$ and $P\rightarrow \infty$, we can obtain the total DoF upper bound as
\begin{align}
d_{total}=\sum\limits_{i=1}^{K} \sum\limits_{j \in {\cal S}_i} d_{i,j} \leq \frac{K^2}{2}\frac{4N}{K^2+\beta^2}=\frac{2K^2N}{K^2+\beta^2}.
\end{align}

Thirdly, we consider the case when $\frac{N}{M} \in \left(\frac{(K^2+(\beta-2)^2)\beta}{K^2+\beta^2}, \beta\right]$. We prove this by contradiction. If $\frac{N}{M}  \in \left(\frac{(K^2+(\beta-2)^2)\beta}{K^2+\beta^2}, \beta\right]$ and $d_{total}^a > \frac{2K^2\beta M}{K^2+\beta ^2}$, where $d_{total}^a$ represents the achivable total DoF, then we increase $N$ to $N_1$ such that $\frac{N_1}{M} =\beta$. Utilizing the antenna deactivation, $d_{total}^a > \frac{2K^2\beta M}{K^2+\beta ^2}=\frac{2K^2 N_1}{K^2+\beta ^2}$ can be achieved. However, this contradicts with that when $\frac{N_1}{M} =\beta$, the DoF upper bound is $\frac{2K^2 N_1}{K^2+\beta ^2}$. Hence, the DoF upper bound of the case when $\frac{N}{M} \in \left(\frac{(K^2+(\beta-2)^2)\beta}{K^2+\beta^2}, \beta\right]$ is $\frac{2K^2\beta M}{K^2+\beta ^2}$.

Finally, we consider the case when $\frac{N}{M} \in \left(\frac{K^2-3K+3}{K-1}, +\infty \right)$. In this case, we notice that the DoF per user could not be larger than $M$. Thus, $KM$ is the DoF upper bound for the case when $\frac{N}{M} \in \big(\frac{K^2-2K+2}{K}, +\infty \big)$.

\balance

\bibliographystyle{IEEEtran}
\bibliography{IEEEabrv,reference}

\begin{IEEEbiographynophoto}{Kangqi Liu}
(S'14) received the B.S. degree in electronic engineering from Shanghai Jiao Tong University, Shanghai, China, in 2013. He is currently pursuing the Ph.D. degree with the Department of Electronic Engineering, Shanghai Jiao Tong University.

His research interests include interference management in wireless networks, network information theory, transmitter and receiver techniques for MIMO systems, and advanced signal processing for wireless cooperative communication.
\end{IEEEbiographynophoto}

\begin{IEEEbiographynophoto}{Meixia Tao}
(S'00--M'04--SM'10) received the B.S. degree in electronic engineering from Fudan University, Shanghai, China, in 1999, and the Ph.D. degree in electrical and electronic engineering from Hong Kong University of Science and Technology in 2003. She is currently a Professor with the Department of Electronic Engineering, Shanghai Jiao Tong University, China. Prior to that, she was a Member of Professional Staff at Hong Kong Applied Science and Technology Research Institute during 2003-2004, and a Teaching Fellow then an Assistant Professor at the Department of Electrical and Computer Engineering, National University of Singapore from 2004 to 2007. Her current research interests include cooperative communications, wireless resource allocation, MIMO techniques, and physical layer security.

Dr. Tao is a member of the Executive Editorial Committee of the \textsc{IEEE Transactions on Wireless Communications} since Jan. 2015. She serves as an Editor for the \textsc{IEEE Transactions on Communications} and the \textsc{IEEE Wireless Communications Letters}. She was on the Editorial Board of the \textsc{IEEE Transactions on Wireless Communications} from 2007 to 2011 and the \textsc{IEEE Communications Letters} from 2009 to 2012. She also served as Guest Editor for \textsc{IEEE Communications Magazine} with feature topic on LTE-Advanced and 4G Wireless Communications in 2012, and Guest Editor for \textsc{EURISAP J WCN} with special issue on Physical Layer Network Coding for Wireless Cooperative Networks in 2010. She has served as the TPC chair of IEEE/CIC ICCC 2014 and as Symposium Co-Chair of IEEE ICC 2015.

Dr. Tao is the recipient of the IEEE Heinrich Hertz Award for Best Communications Letters in 2013, the IEEE ComSoc Asia-Pacific Outstanding Young Researcher Award in 2009, and the International Conference on Wireless Communications and Signal Processing (WCSP) Best Paper Award in 2012.

\end{IEEEbiographynophoto}

\end{document}